\documentclass[jgrga]{agu2001}
\usepackage{lineno}
\usepackage{graphicx,color}

\renewcommand{\aap}{{\it Astron. Astrophys.}}

\renewcommand{\apjl}{   {\it Astrophys. J. Lett.}}

\authorrunninghead{DASSO ET AL.}
\titlerunninghead{TRACKING TWO CONSECUTIVE MAGNETIC CLOUDS}

\begin{document}
\pdfoutput=1
\setkeys{Gin}{draft=false}
\title{Linking two consecutive non-merging magnetic clouds with their solar sources}

\authors{
S. Dasso \altaffilmark{1,2}, C.H. Mandrini \altaffilmark{1}, B. Schmieder \altaffilmark{3},
H. Cremades \altaffilmark{4}, C. Cid \altaffilmark{5}, Y. Cerrato \altaffilmark{5},
E. Saiz \altaffilmark{5}, P. D\'emoulin \altaffilmark{3}, A.N. Zhukov \altaffilmark{6,7},
L. Rodriguez \altaffilmark{6}, A. Aran \altaffilmark{8}, M. Menvielle \altaffilmark{9}, and
S. Poedts \altaffilmark{10}}

\altaffiltext{1}
{Instituto de Astronom\'{i}a y F\'{i}sica del Espacio (IAFE), CONICET-UBA,
Buenos Aires, Argentina}
\altaffiltext{2} {Departamento de F\'{i}sica, FCEN-UBA, Buenos Aires,
Argentina}
\altaffiltext{3} {Observatoire de Paris, LESIA, UMR 8109 (CNRS), Meudon,
France}
\altaffiltext{4} {Universidad Tecnol\'ogica Nacional, Facultad Regional
Mendoza, Argentina}
\altaffiltext{5} {Universidad de Alcal\'a, Alcal\'a de Henares, Spain}
\altaffiltext{6} {Solar-Terrestrial Center of Excellence - SIDC, Royal
Observatory of Belgium, Brussels, Belgium}
\altaffiltext{7} {Skobeltsyn Institute of Nuclear Physics, Moscow State
University, Moscow, Russia}
\altaffiltext{8} {Universitat de Barcelona, Barcelona, Spain}
\altaffiltext{9} {Centre d\'etude des Environnements Terrestre et
Plan\'etaires, V\'elizy, France}
\altaffiltext{10} {K.U. Leuven, Leuven, Belgium}

\begin{abstract}
  On May 15, 2005, a huge interplanetary coronal mass ejection (ICME) was
observed near Earth. It triggered one of the most intense geomagnetic storms
of solar cycle 23
($Dst_{peak}$ = -263~nT).
This structure has been associated with the two-ribbon flare,
filament eruption and CME originating in active region (AR)~10759
(NOAA number).
We analyze here the sequence of events, from solar wind measurements (at
1~AU) and back to the Sun, to understand the origin and evolution of this
geoeffective ICME.
From a detailed observational study of in situ magnetic field
observations and plasma parameters in the interplanetary (IP) medium, and
the use of appropriate models, we propose an alternative interpretation of
the IP observations, different to those discussed in previous studies. In
our view, the IP structure is formed by
two extremely close consecutive magnetic clouds (MCs) that preserve
their identity during their propagation through the interplanetary medium.
Consequently, we identify two solar events in H$\alpha$ and EUV
which occurred
in the source region of the MCs.
The timing between solar and IP events, as well as
the orientation of the MC axes and their associated solar arcades are
in good agreement.
Additionally, interplanetary radio type~II observations allow the tracking
of the multiple structure through inner heliosphere and to pin down the
interaction region to be located midway between the Sun and the Earth.
The chain of observations from the photosphere to interplanetary space is in
agreement with this scenario.
Our analysis allows the detection of the solar sources of the transients and explains
the extremely fast changes of the solar wind due to the transport of two
attached ---though non-merging--- MCs which affect the magnetosphere.
\end{abstract}

\begin{article}

\section{Introduction}
\label{Introduction}

  Coronal mass ejections (CMEs) remove plasma and magnetic field from the Sun
expelling them into interplanetary space. When they are detected in the
interplanetary (IP) medium, they are called  interplanetary coronal mass
ejections (ICMEs).
The transit time of an ICME from the Sun to 1~AU is typically in the interval
1 to 5 days
\citep{Gopalswamy00,Gopalswamy01,Rust05}.
A subset of ICMEs, called magnetic clouds (MCs), are characterized by an
enhanced magnetic field strength, a smooth and large rotation of the magnetic
field vector, and low proton temperature \citep{Burlaga81}.
Fast and large magnetic clouds are mostly observed at 1~AU in the
declining phase of a solar cycle, as it happened in the last one
during 2003-2005 \citep[e.g., ][]{Culhane07}.

  Interplanetary type~II bursts permit the tracking of fast ICMEs through the analysis
of radio emission frequency drift \citep{Reiner98,Reiner07,Hoang07}.
Assuming the type~II emission to be produced at the fundamental or second harmonic
of the local plasma frequency, and with help of a heliospheric density model,
the radio frequency can be converted to radial distance. This method has
still limitations and allows for possible different interpretations due to the
patchiness and the frequency range of the observed radio emissions.
The evolution of an ICME can also be followed along its journey
using a combination of solar and IP observations together
with the results of numerical simulations \citep[e.g., ][]{Wu99}.

  Coronal mass ejections are frequently associated with filament
eruptions. The directions of the MC axes are found to be roughly
aligned with the disappearing filaments \citep{Bothmer94,Bothmer98},
preserving their helicity sign.
This result has been found for some individual cases by
\cite{Marubashi97,Yurchyshyn01,Ruzmaikin03,Yurchyshyn05,Rodriguez08}.
Some quantitative (quantifying magnetic fluxes and helicities)
studies of MCs and their solar sources have been also done
\citep{Mandrini05,Luoni05,Longcope07}.
But again, it is not so easy to quantify this association
and further developments are needed
to really understand the
fundamental transport mechanisms and interaction with
the ambient solar wind (see, e.g., the review by \cite{Demoulin08a}).

 Furthermore, when multiple CMEs are expelled from the Sun,
they can be merged leading to the so-called ``CME cannibalism''
\citep{Gopalswamy01b}, mainly as a consequence of magnetic reconnection.
The interaction of two CMEs in favorable conditions for reconnection
has been studied by \cite{Wang05c}.
However, from a theoretical point of view, two original structures can be
preserved with an appropriate orientation of the ejected flux rope
(yielding almost parallel interacting magnetic fields). For example, in the
MHD simulations of \cite{Xiong07} the two interacting flux ropes preserved
their identity while evolving in the IP medium until they reach the Earth
environment or even beyond.

  In the magnetohydrodynamic (MHD) framework,
a magnetic cloud configuration in equilibrium can be obtained
from the balance between the magnetic Lorentz force and the
plasma pressure gradient. Several magnetostatic models have been
used to describe the configuration of MCs.
Frequently, the magnetic field of MCs has been modeled by
the so-called Lundquist's model \citep{Lundquist50}, which considers a
static and axially symmetric (cylindrical) linear force-free
magnetic configuration neglecting the plasma pressure \citep[e.g., ][]
{Goldstein83,Burlaga88,Lepping90,Lynch03,Dasso05,Leitner07,Xiong07}.
The azimuthal and axial field components of the flux rope
in this classical configuration are defined
by $B_\phi=B_0 J_1(\alpha r)$ and $B_z=B_0 J_0(\alpha r)$,
where $r$ is the distance to the MC axis, $J_n$ is the Bessel function
of the first kind of order $n$, $B_0$ is the strength
of the field at the cloud axis, and $\alpha$ is a constant associated
with the twist of the magnetic field lines.
Some other refined models have also been proposed to describe the magnetic
structure of clouds \citep[e.g., ][]{Hu01,Vandas02,Cid02}; in particular, some
of them consider an elliptical shape \citep{Hidalgo02,Hidalgo03}
that allows the description of possible distortions of the structures.

   On May 15-17, 2005, the strongly southward interplanetary field
(above 40~nT) and the high solar wind velocity (close to 1000~km~s$^{-1}$),
observed by the Advanced Composition Explorer (ACE), are the
cause of a super geomagnetic storm with a depression of the $Dst$ index
reaching -263~nT. The IP structure has the characteristics of a
MC. \cite{Yurchyshyn06} associated this structure with the two-ribbon M8.0
X-ray class flare on May 13 at 16:32~UT in AR~10759, accompanied by a filament
eruption and CME.
These authors identified a single MC, starting at $\sim$ 00:00~UT on
May 15 and ending at $\sim$ 09:00~UT on May 16. An event with different
boundaries, much smaller than the one described by Yurchyshyn et al.,
was identified by R. Lepping (start on May 15 at 05:42~UT and end
on May 15 at 22:18~UT). The event was qualified with intermediate fitting
quality (quality 2) and had a large impact parameter (catalog at
http://lepmfi.gsfc.nasa.gov\-/mfi\-/mag\_cloud\_S1.html).

   The solar activity evolution along May 13 has been well described by
\cite{Yurchyshyn06} and, more recently, by \cite{Liu07}. Both papers refer
mainly to the M8.0 event and conclude that the eruption of a large
sigmoidal structure launches the CME observed by the Large Angle and
Spectroscopic Coronagraph \citep[LASCO,][]{Brueckner95} at 17:22 UT.

In this paper, we present a detailed
analysis of the solar wind conditions at the Lagrangian L1 point using in situ
observations (magnetic field and bulk plasma properties) and
propose an alternative interpretation of the IP observations. In our view,
the observed ICME is in fact formed by two MCs.
Since a single solar event cannot explain the arrival of two MCs
at 1~AU, we revisit the evolution and activity of AR~10759 and other
regions present in the solar disk. We start searching in the Sun for two possible
sources of the clouds, even earlier than May 13.
We identify a previous candidate event in H$\alpha$ and EUV data same day at 12:54~UT
in AR~10759, which was classified as a C1.5 flare in GOES. These
two events, the one at 12:54~UT and the one at 16:32~UT, give rise
to two two-ribbon flares along different portions of the AR magnetic
inversion line. The orientations of the magnetic fields associated to both solar events
are in good agreement with the fields observed in their associated MCs.

In Section~\ref{solar-wind} we analyze the solar wind conditions near Earth;
in particular, we derive the orientation of the axis for the two MCs by
fitting a model to the observations. In Section~\ref{solar}, we revisit the
solar events using ground based and satellite data. In
Section~\ref{radio-obs} we present radio type~II
remote observations, which let us track the multiple structure through the inner heliosphere.
Finally, in Section~\ref{physical_scenario_discu}, we
describe the most plausible physical scenario behind the chain of events
from May~13 to 17, 2005, and we give our conclusions.

\section{A Large Interplanetary Coronal Mass Ejection: May 15-17.}
\label{solar-wind}

\subsection{In situ plasma and magnetic field observations}
\label{insitu-icme-obs}

   The solar wind data were obtained by ACE, located at the
Lagrangian point L1.
The magnetic field observations come from the
{\it Magnetic Fields Experiment} (MAG)
\citep{Smith98} and the plasma data from the {\it Solar Wind Electron
Proton Alpha Monitor} (SWEPAM) \citep{McComas98b}.

\begin{figure*}
\centering
\includegraphics[width=39pc]{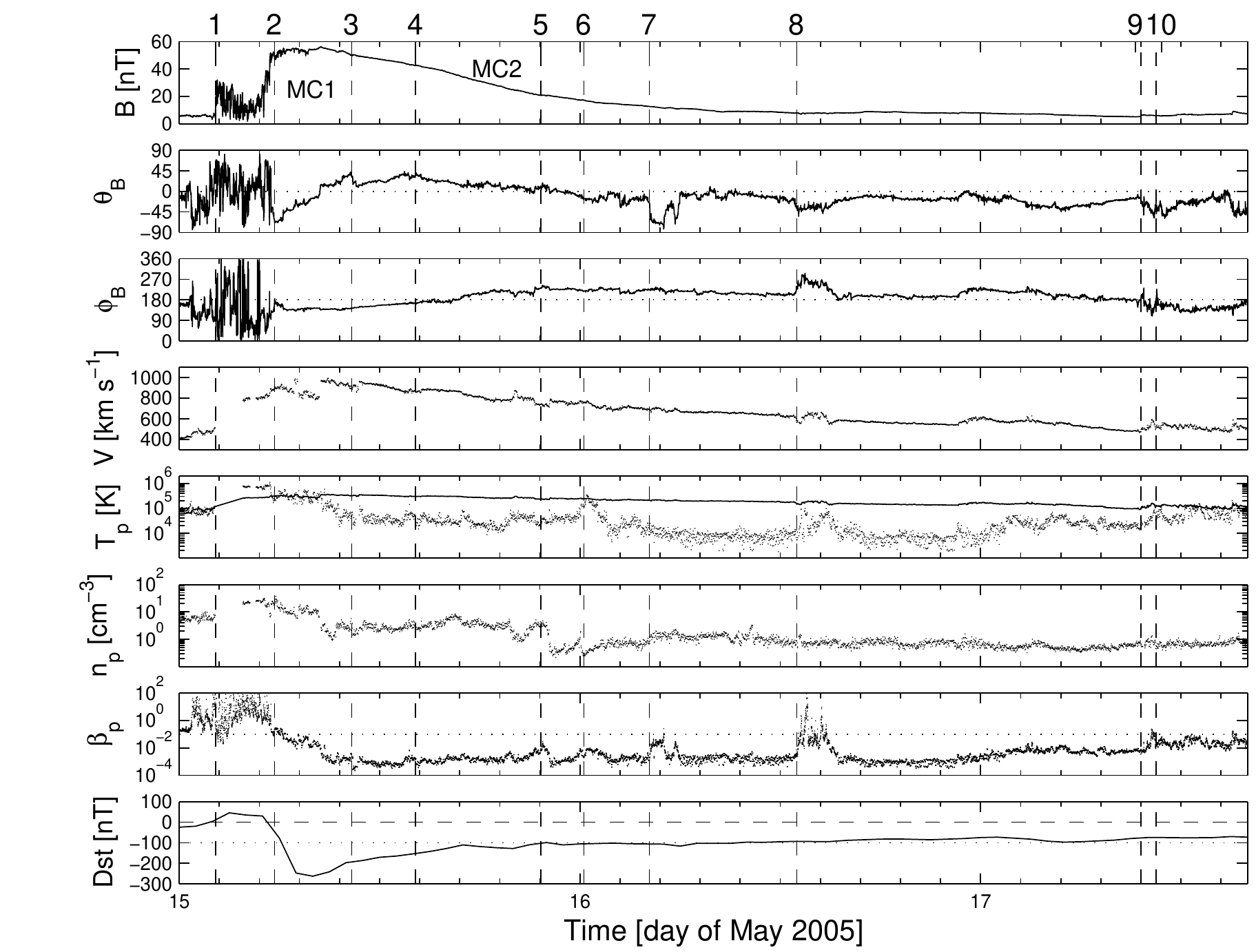}
\caption{In situ (at L1) plasma and magnetic field observations (ACE) of the
ICME on May 2005. From top to bottom: absolute value
of the magnetic field ($B=|\vec{B}|$), magnetic field vector
orientation (GSE): latitude ($\theta_B$) and longitude ($\phi_B$), bulk
velocity ($V$), expected (continuous line)
and observed (dots) proton temperature
($T_p$), proton density ($n_p$), proton plasma beta ($\beta_p$),
and $Dst$ index. Vertical lines mark different interfaces
(see Section~\ref{insitu-icme-obs} for a description and
Table~\ref{tabla_ticks} for timings). Horizontal dotted lines
in $\theta_B$, $\phi_B$, $\beta_p$, and $Dst$ mark values at 0, 180, 0.1,
and -100 as a reference, respectively.}
\label{fig-ICME-VB}
\end{figure*}

   Figure~\ref{fig-ICME-VB} shows the magnetic field (in the
Geocentric Solar Ecliptic, GSE, system) and plasma conditions
in the solar wind for the long-lasting studied event
(00:00~UT on May 15 to 16:00~UT on May 17, 2005),
and the consequent magnetospheric activity ($Dst$ index).

  The data in Figure~\ref{fig-ICME-VB} show several features consistent with
the presence of
ICMEs/MCs: low proton plasma beta, proton temperature lower than
expected ($T_{ex}$) for
a typical solar wind at a given observed bulk velocity \citep{Lopez87},
an almost linear velocity profile,
and a smoothly varying magnetic field orientation of high intensity.
In particular, from vertical dashed line '1' to '10'
(for timings of tick numbers see next paragraphs and Table~\ref{tabla_ticks})
the magnetic field intensity is strongly enhanced, although the scale needed
to show the whole event in one plot does not allow to observe this enhancement
at the end of the time interval.
The temporal length of the event is quite large, so that it corresponds to a very
extended region, $\sim 0.93$~AU, one of the largest ICMEs ever observed
(see, e.g., Table 1 of \cite{Liu05} and Figure 1 of \cite{Liu06b}).
At '10', the magnetic field recovers its background value of $\sim 5$~nT.

From a comparative study between magnetic clouds and complex ejecta,
\cite{Burlaga01} found that while most of the clouds were associated with
single solar sources, nearly all the complex ejecta could have had multiple sources.
Evidence was also found indicating that long duration complex merged interaction regions
(with radial extents of $\sim 0.7$ AUs) can be produced by the interaction
of two or more CMEs/MCs/shocks \citep{Burlaga01,Burlaga03}.

  In front of the ICME, from 00:00~UT to 02:11~UT on May 15 (labeled as '1'), the
solar wind presents  typical conditions with a value of $B \sim$ 6~nT, an
increasing velocity profile (starting at $\sim$ 400~km~s$^{-1}$ and reaching
$\sim$ 500~km~s$^{-1}$), observed proton temperature similar to that expected
one ($T_{ex}$) for a typical solar at same velocity, an increasing proton
density profile from 3 to 10 protons per cm$^{3}$, high values of proton plasma $\beta_p$
(reaching $\sim$ 10-100), and values of the $Dst$ index ($Dst \sim 0$)
corresponding to relatively quiet ring current conditions.

\begin{table}
\caption{Timings (dd,hh:mm~UT) and substructures inside the ICME of May 15-17,
2005,their boundaries are identified with numbers in Figure~\ref{fig-ICME-VB}.}
\begin{tabular}{ccc}
\hline
Tick number  & Timing & Substructure \\
\hline
1 & 15,02:11 &      \\
  &          &sheath\\
2 & 15,05:42 &      \\
  &          & MC1  \\
3 & 15,10:20 &      \\
  &          & back1\\
4 & 15,14:10 &      \\
  &          & MC2  \\
5 & 15,21:40 &      \\
  &          & MC2  \\
6 & 16,00:15 &      \\
  &          & MC2  \\
7 & 16,04:10 &      \\
  &          & back2\\
8 & 16,13:00 &      \\
  &          & back2\\
9 & 17,09:37 &      \\
  &          & back2\\
10& 17,10:30 &      \\
\hline
\end{tabular}
\label{tabla_ticks}
\end{table}

  Observations at '1' suggest the existence of a strong leading edge shock
(not fully confirmed from observations of $V$ because of a gap in plasma data from
02:11~UT to 03:50~UT on May 15). Just behind the shock,
from '1' to '2',  a typical ICME sheath is present
(e.g., high level of fluctuations in $\vec{B}$, enhanced B strength, high
mass density,
high $\beta_p$).

  A structure with a very high $B$ (50-60~nT), one of the highest values ever
observed in the solar wind at 1~AU, is found between '2' and '3'.
In this range of time, a large scale coherent rotation of the magnetic field
vector is present,
with $\vec{B}$ going from south to north (see $\theta_B$ panel).
This is a left handed SEN flux rope in the classification of \cite{Bothmer98}.
Note the presence of a sudden change of the sense of rotation together
with a magnetic discontinuity (MD) at '3' (clearly observed in
$\theta_B$).
A current sheet (i.e., a discontinuity in the observed time series of the
magnetic field vector) is expected
to be present at the interface that separates two magnetic regions with
different magnetic connectivity, i.e., with different magnetic stress.
Thus, we interprete the discontinuity at '3' as the signature
of the end of a first substructure inside the large ICME.
This substructure '2-3' has also specific physical properties:
there are no signatures of expansion (the profile of $V$ does not show a significant
slope), and the observed temperature decreases in time (hotter near the sheath and
colder near substructure '3-4').
As a consequence of the decreasing $T_p$ profile, values of
$\beta_p$ also decrease, being $\beta_p \sim 10^{-1}$ near '2' and
reaching values as low
as $\beta_p \sim 10^{-3}$ near '3'. Thus, the substructure '2-3' presents
some MC signatures (low $\beta_p$, coherent rotation of the magnetic field
vector) but not all ICMEs features (e.g., $T_p$ is not significantly lower
than $T_{ex}$). This MC presents a short temporal duration with
a spatial size along the Sun-Earth direction of $\sim$0.09 AU
($\sim$4 hs multiplied by $\sim$900~km~s$^{-1}$).
We interpret the short duration of the first MC as due to the
compression made by MC2 at its trailing edge; it is expected that the
expansion rate (and the consequent cooling) of the first cloud
diminished during part of its journey, where the second
MC is pushing it from behind \citep{Wang03}, consistently with
observations ($T_p \sim T_{ex}$).

After the MD at '3', $\theta_B$ continues
increasing up to '4', where there is another (weaker, but significant) MD.
The region from '3' to '4' shows the characteristics of the back of flux ropes,
previously found by \cite{Dasso06} in a significantly
large non-expanding magnetic cloud (Oct. 1995) and by \cite{Dasso07} in
another huge magnetic cloud (Nov. 2004) in strong expansion.
The formation of this back feature is a consequence of
previous magnetic reconnection between the front of the flux rope and its
environment;
magnetic flux is removed from the flux rope front (the front is peeled) while
its counterpart in
the rear part still remains (Figure 6 of \cite{Dasso06}).
The formation of a back in fast ICMEs is also supported by numerical
simulations (see Figure 4 of \cite{Wu05}).

At the difference of previously studied MC-backs,
the region from '3' to '4' is compressed by the second MC.
This region of interaction does not have the same characteristics as
corotating interaction regions \citep[CIRs, e.g.][]{Pizzo94} or typical merged interaction regions
\citep[MIRs, e.g.][and references there in]{Burlaga93}. CIRs are due to a fast SW overtaking
a slow SW, and they develop at larger distance from the Sun ($\geq 10$~AU) than MIRs.
The physics involved in the region between two interacting MCs is expected to be different, since
MCs have a moderate spatial extension and are structured by the magnetic field.
It implies that the compressed plasma can be evacuated on the sides, and that the second MC is able
to accelerate fully the first one.  Without a significant reconnected flux between the two MCs
(when the magnetic fields are nearly parallel) and after a transient period of time from the interaction,
the two MCs are expected to travel together (a situation fully different than in the case
of a fast SW overtaking a slow SW).

The next substructure that can be identified is '4-7'.
It is a very huge region ($\sim$ 0.3~AU) that presents
a low variance of $\vec{B}$ and clear signatures of an expanding MC,
with a very strong magnetic field that rotates coherently (from north-east to
south-west). This region shows an almost linear decay of $B$
(from $B \sim 60$ to $B \sim 15$~nT) consistent with the observed expansion,
with a linear $V$ profile from 900 to 700~km~s$^{-1}$ during $\sim 14$ hours
(equivalent to $\sim$ 0.3~AU), a typical expansion rate observed in MCs \citep{Demoulin08b}.
It also presents values of $T_p$ significantly lower than $T_{ex}$,
low proton density ($n_p \sim 3$ cm$^{-3}$), and very low values of
$\beta_p$ ($\sim 10^{-3}$). All these signatures discard other well-known non-MCs
structures (as e.g. corotating interaction regions).

Inside region '4-7', we also remark '5' and '6'.
At '5' a strong decrease of $n_p$ starts and there are small variations in the
large scale trend of $\theta_B$, $\phi_B$, $B$, and $V$.
At '6', $\beta_p$, $n_p$, and $T_p$ have peaks (here $T_p$ reaches $T_{ex}$).
Thus, it is a priori unclear where to set the end
of the second flux rope, we can distinguish three possible ends:
'5', '6', or '7'.
Finally, we set it at '7' because, at this position, a very strong MD is found
and the coherence of $\vec{B}$ is lost.
The presence of two flux ropes inside this ICME is also supported by applying
the Grad-Shafranov technique \citep{Hu02} to the plasma and magnetic field
data at L1 (M{\"o}stl, C., 2007, private communication).

An expanding structure with ICME signatures (e.g., low proton temperature,
low $\beta_p$, decreasing $V$ and $B$ profiles) still remains after '7'.
At position '8', there is another MD
(observed mainly in $\theta_B$ and $\phi_B$) together with a change in the
decay rate of $B$. This MD is associated with a sudden increase of $T_p$
(and consequently of $\beta_p$, because $n_p$ remains approximately constant),
while the velocity profile still presents roughly the same slope
(i.e. the same expansion rate).
Even when the expansion signatures (in both, $V$ and $B$) and the
coherence of $\vec{B}$ end at '9',
low $\beta_p$ and $T_p<T_{ex}$ remain along one hour, until '10'.  Here $B$
recovers values as low as
those found in the typical solar wind, $T_p$ recovers the expected values
($T_p \sim T_{ex}$),
the coherence in the rotation of $\vec{B}$ is lost, and the level of
fluctuations
(e.g., in $\vec{B}$, see panels $\theta_B$ and $\phi_B$) start to increase
significantly.
This large region, from '7' to '10', shows the characteristics of an extended
back feature belonging to the second flux rope.

The last panel of Figure~\ref{fig-ICME-VB} shows the geomagnetic
effect of the ICME,
as monitored by the $Dst$ index (preliminary index downloaded from the
OMNI data base). From '2' the intensity of the storm
increases, because of the high intensity of the driven electric field
($V B_s \sim 5\times 10^4$~nT km~s$^{-1}$,
with $B_s$ being the southern component of interplanetary magnetic field).

The storm starts its recovery phase at $\sim$ 08:00~UT, when the
magnetospheric response is the strongest (reaching $Dst = -263$~nT);
we notice a significant change in the decay rate after '3'.
Later, $Dst$ remains $ < -100$~nT during part of the recovery phase
(until '5').
Assuming a pure decay, decay times of $\tau \sim$ 5 hours and $\tau
\sim$ 17 hours
are found for the first (from the peak of the storm to '3') and second stage
(beyond '3'), respectively.
These two values are beyond the lower and close to the upper border of the
typical range obtained for the ring current decay time, $\tau = 14 \pm 4$ hours,
by  \cite{Dasso02}. Larger decay times for the recovery phase have been
associated with the presence of multiple IP structures near Earth
\citep{Xie06}.

\begin{figure}
\centering
\includegraphics[width=\linewidth]{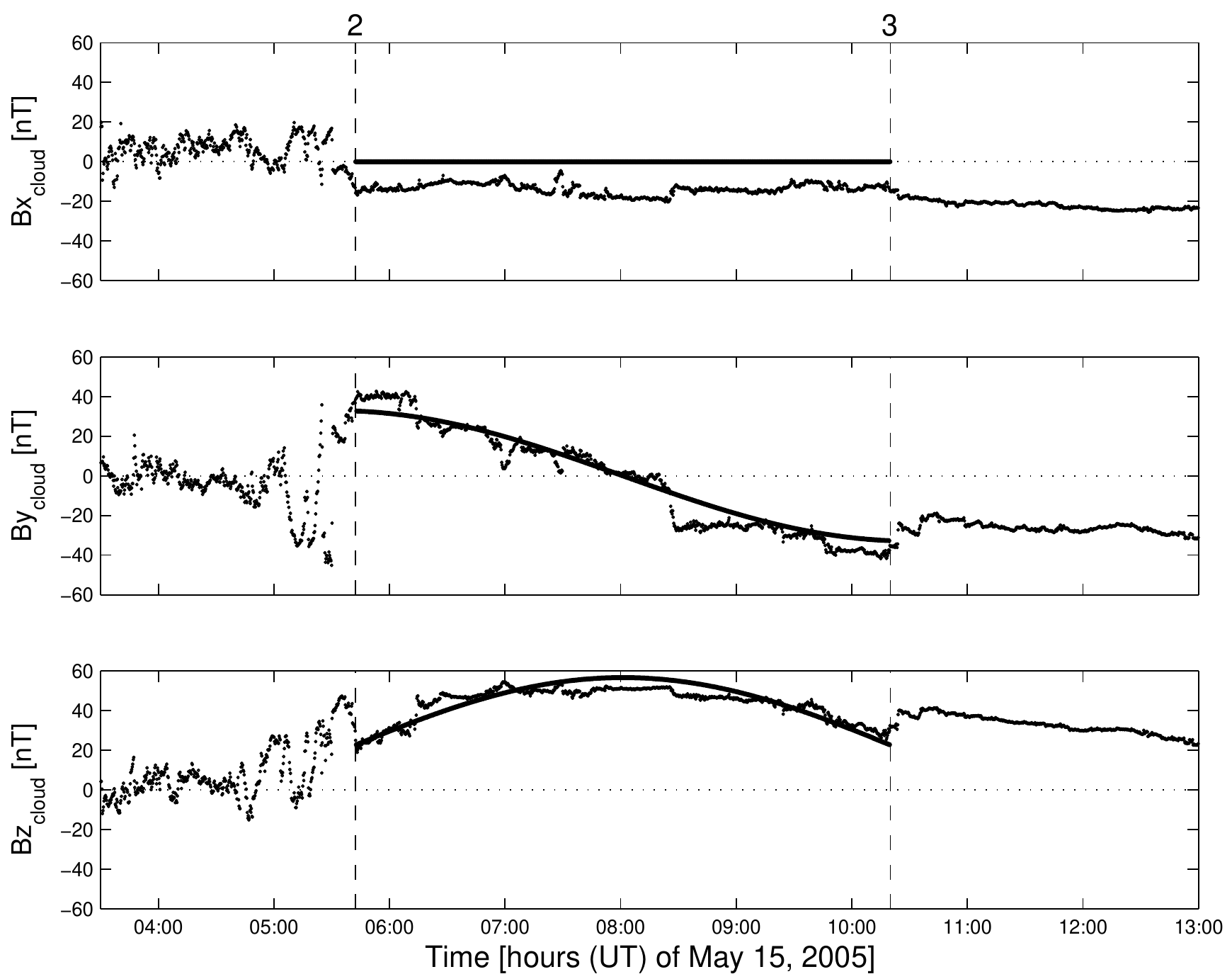}
\caption{ACE observations (dots) of the magnetic field vector
components in the cloud frame (see Section~\ref{mcs-models-23})
for the first flux rope (observations between '2' and '3'). From upper
to lower panels: $B_{x, \rm cloud}$, $B_{y, \rm cloud}$,
and $B_{z, \rm cloud}$.
The solid lines correspond to the fitted curves obtained using the Lundquist's model.}
 \label{fig-mv-Lundfit-1stMC}
\end{figure}

\subsection{First Magnetic Cloud}
\label{mcs-models-23}

In this section, we model the magnetic structure following
the sheath (i.e., the data between '2' and '3') that we call the first cloud
(MC1).

  To better understand the MC properties,
we define a coordinate system linked to the cloud
in which $\hat{z}_{\rm cloud}$ is oriented along the cloud axis
(with $B_{z, \rm cloud}>0$ at the cloud axis).
Since the speed direction of a cloud is mainly aligned with
the Sun-Earth direction and is much larger than the spacecraft speed,
we assume a rectilinear spacecraft trajectory in the cloud frame.
This trajectory defines a direction $\hat{d}$ (pointing toward the Sun).
Then, we define $\hat{y}_{\rm cloud}$ in the
direction $\hat{z}_{\rm cloud} \times \hat{d}$ and finally $\hat{x}_{\rm
cloud}$ completes the right-handed orthonormal base ($\hat{x}_{\rm
cloud},\hat{y}_{\rm cloud},\hat{z}_{\rm cloud}$).

  We define the axis latitude
angle ($\theta$) as the angle between  the cloud axis and
the ecliptic plane, and the axis longitude angle ($\varphi$) as the one
between the projection of the cloud axis on the ecliptic
plane and the Earth-Sun direction ($\hat{x}_{GSE}$)
measured counterclockwise \citep[see][]{Dasso06}.
We also define the impact parameter, $p$, as the minimum
distance from the spacecraft to the cloud axis.

  The local coordinate system is especially useful when $p$ is small
compared to the MC radius ($R$).  In particular, for $p=0$ and a cloud
described by a cylindrical magnetic configuration $\vec{B}(r) = B_z(r)
\hat{z} + B_\phi(r) \hat{\phi}$, we have $\hat{x}_{\rm cloud} =
\hat{r}$ and $\hat{y}_{\rm cloud} = \hat{\phi}$ when the spacecraft
leaves the cloud.
In this particular case, the magnetic field data will
show: $B_{x, \rm cloud}=0$, a large and coherent variation of $B_{y,
\rm cloud}$ (with a change of sign), and an intermediate and coherent
variation of $B_{z, \rm cloud}$, from low values at one cloud edge,
achieving the maximum value at its axis, and returning to low values at
the other edge.

The minimum variance (MV) method \citep{Sonnerup67} has been used to
estimate the orientation of MCs [see e.g., \citealt{Bothmer98},
\citealt{Lepping90}, \citealt{Farrugia99}, \citealt{Dasso03},
\citealt{Gulisano05}].
In particular, using a synthetic set of
ideal cylindrical clouds, \cite{Gulisano07}
have shown that the application of the MV technique to the
normalized observed time series of the magnetic field
($\vec{B}(t)/|\vec{B}(t)|$) can provide very good estimations
of the cloud axis (when $0 \leq p/R \leq 0.7$). This is feasible
because when $\vec{B}(t)$ is normalized,
the information of the rotation of the field vector
is not mixed with possible changes in its absolute value
during the observations.

We apply the normalized MV method to the
observations between '2' and '3'. We find a left handed flux rope,
oriented such that $\theta_{MV}$=-15$^\circ$ and
$\varphi_{MV}$=125$^\circ$, consistent with the SEN orientation
determined in Section~\ref{insitu-icme-obs}.
From this orientation and the mean bulk speed (894~km~s$^{-1}$ for
this range), we estimate the cloud size perpendicular to its axis as
$\sim 0.08$~AU.

In the MC frame (see Figure~\ref{fig-mv-Lundfit-1stMC}), the magnetic field
has the typical shape observed in clouds. The almost constant
$B_{x, \rm cloud}$ profile (with a mean value
$<B_{x, \rm cloud}> \sim -11.5$~nT), indicates that $p$ is not zero.
Moreover, a rough estimation of $p$ can achieved using the
expression found by \cite{Gulisano07}
($p/R \sim \sqrt{(<B_{x, \rm cloud}>/B_0)/1.6}$), which gives
$p \sim (0.3-0.4)R$.

The MC borders set as '2' and  '3' are in agreement with the expected
conservation of magnetic flux across a plane perpendicular to $\hat{x}_{cloud}$
\citep{Dasso06}, i.e., similar areas below and above the
curve $B_{y, \rm cloud}$ which equals before and after the MC center.
These boundaries are also in agreement with
the expected MD (current sheets) at the interfaces
between two structures with different connectivity
(as the boundaries of a flux rope forming a MC).

The MC axis, with $\varphi_{MV}=125^\circ$, is dominantly pointing
towards $\hat{y}_{GSE}$ with a significant contribution
toward -$\hat{x}_{GSE}$. This is consistent with the spacecraft passing
through the right (west) leg of the flux rope and,
thus, with the MC apex located toward the left of L1
(observed from Earth to the Sun, with north upward). From the sign of the
observed $<B_{x, \rm cloud}>$, the cloud axis is toward the south of the
ecliptic.

  Fixing the orientation provided by the MV method,
we fit the free parameters of Lundquist's model
(see Section~\ref{Introduction} for the meaning of the free parameters)
and obtain $B_0=57$~nT and $\alpha=-40$~AU$^{-1}$.
Solid lines in panels $B_{y, \rm cloud}$
and $B_{z, \rm cloud}$ in Figure~\ref{fig-mv-Lundfit-1stMC}
show the curves obtained from the fitting, which
are in a very good agreement with the observations (dots).

  To validate our previous results, we perform a simultaneous fitting (SF)
of the geometrical and physical free parameters using the same procedure and
numerical code as \cite{Dasso06}. We obtain
$\theta_{\rm SF}$=-12$^\circ$, $\varphi_{\rm SF}$=129$^\circ$,
$R_{\rm SF}=0.04$~AU, $p/R_{\rm SF}=0.2$, $B_{0, \rm SF}=59$~nT,
$\alpha_{\rm SF}=-42$~AU$^{-1}$. The agreement between the results of the SF
and those from the MV method followed by the fitting of the physical cloud
parameters is well within the precision of the methods
(the differences are only $\sim 4^\circ$ in the orientation,
5\% in the radius and 3\% in $B_{0}$).
Due to the radial propagation of the solar wind and due to a possible
meridional stratification of the solar wind properties, distortions from a
cylindrical cross section are expected \citep[see, e.g., ][]{Liu06a}.
From a multispacecraft analysis of the first MC observed by STEREO,
an oblated transverse size with the major axis perpendicular to the Sun-Earth
direction was found \citep{Liu08}.
The good match between the observations and the cylindrical model
(Figure~\ref{fig-mv-Lundfit-1stMC})
indicates that this flux rope could have an almost cylindrical
configuration (a significant ellipticity would have consequences on
the magnetic field profile, see \citet{Vandas03}). Therefore, we conclude that
the main characteristics of MC1 are well determined.

\begin{figure}
\centering
\includegraphics[width=\linewidth,clip=,bb=0 155 584 721]{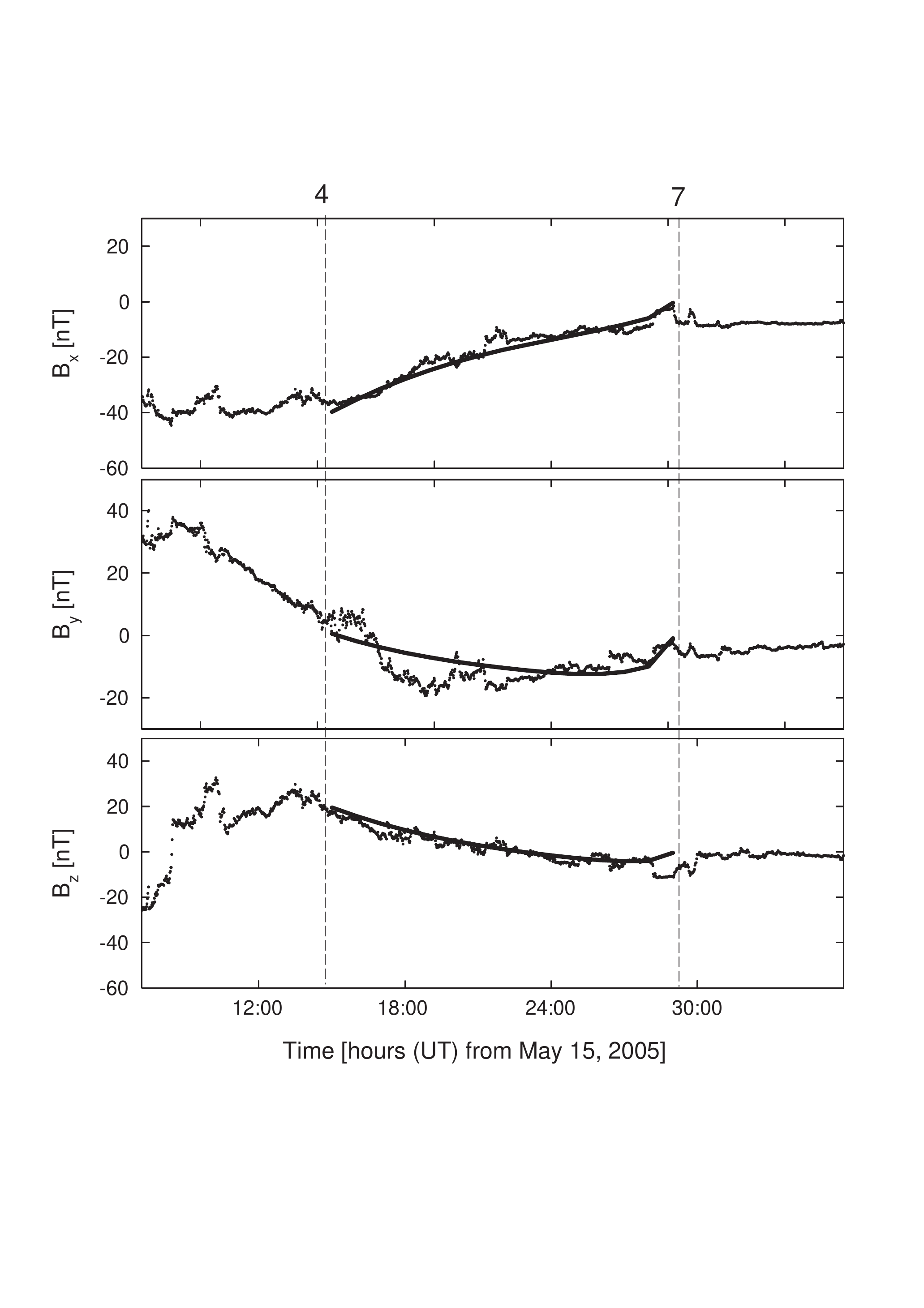}
\caption{Magnetic field vector in GSE components from ACE data (dots)
for the second MC (between '4' and '7') and fitted curve (solid line)
using the model of \cite{Hidalgo02b}.}
 \label{fig-ellipical-2ndMC}
\end{figure}

\subsection{Second Magnetic Cloud}
\label{mcs-models-47}

In this section we model the second MC (MC2) observed from '4' to '7'.
A decreasing velocity profile (see Section~\ref{insitu-icme-obs}) indicates
that the cloud is in expansion. The sudden change of $\theta_B$ at '7'
is the most significant of all changes observed in the field after '4';
furthermore, after this discontinuity, the expected coherence
of $\vec{B}$ is lost. Thus, as previously discussed in
Section~\ref{insitu-icme-obs}, we choose '7' as the rear boundary of MC2.

   The use of the MV method for this second MC does not
provide meaningful results, since the impact parameter is very large as
indicated by the low rotation of the magnetic field vector
(Figure~\ref{fig-ICME-VB}) and also because
$B_{x,\rm cloud}$ is the largest field component in the cloud frame
[not shown].
Moreover, there is significant expansion, thus normalizing the field is
not enough to fully remove this effect.
The SF to Lundquist solution, even with a normalized field,
cannot provide in this case a reliable result for the same reasons.
When the impact parameter is low and the boundaries of the flux rope
are well determined, flux rope modelling generally
provides a good representation of the magnetic field configuration of a MC
\citep{Riley04c}. However, models need to be tested using simultaneous
observations of different parts of the same flux rope,
e.g. as recently done using STEREO observations \citep{Liu08}
confirming the flux rope geometry of the studied event.
Since simultaneous observations from spacecraft with a significant separation
are not available for this event, we apply a different model and method for
MC2, described below. This model has more freedom in comparison with
Lundquist's model used before, such as the oblateness of the cloud cross
section \citep{Liu06a}.

  Observations of some expanding MCs traveling in the solar wind are
consistent with cylindrical expansion \citep[e.g., ][]{Nakwacki05,Dasso07,Nakwacki08}.
However, we can anticipate distortions from the cylindrical shape
for MC2 because of the presence of MC1 in its front.
Therefore, we compare observations with the model by \cite{Hidalgo02b},
which considers a magnetic field with an elliptical cross section.
This model has eight free parameters: the magnetic field strength at the
cloud axis,
the latitude ($\theta$) and longitude ($\varphi$) of the cloud axis, the
impact parameter
($p$), the orientation of the elliptical cross section relative to the
spacecraft path ($\zeta$), a
parameter related to the eccentricity of the cross section ($\eta$), and two
other
parameters related to the plasma current density.

  The fitted curves using the model of \cite{Hidalgo02b} are in very good
agreement with observations
(Figure~\ref{fig-ellipical-2ndMC}). The model corresponds to a left handed
magnetic cloud with
$\theta=(55\pm10)^\circ$ and $\varphi=(120\pm10)^\circ$.

\section{Solar clues for two source events}
\label{solar}

\begin{figure}
\centering
\includegraphics[width=1.1\linewidth]{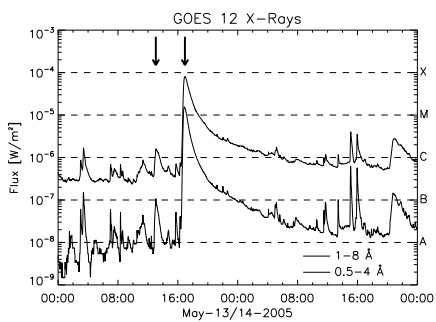}
\caption{GOES lightcurve from May 13 to 14. The flares related
to the eruptions we consider to be the sources of the two MCs
(see Sections~\ref{mcs-models-23} and~\ref{mcs-models-47})
are pointed by arrows at peak intensity
(13:04~UT and 16:57~UT for the first and second flare, respectively).}
 \label{fig-goes}
\end{figure}

\subsection{The data}
\label{solar-obs}

  In this section, we analyze solar data from the photosphere to the
upper corona in search of two candidate solar events that could be the sources of the
two MCs described in previous sections.

    The photospheric magnetic field evolution
is analyzed using observations from the Michelson Doppler Imager
\citep[MDI][]{Scherrer95}, on board the Solar and Heliospheric
Observatory (SOHO), which measures the line of sight magnetic field at the
photosphere.  These data are the average of 5 magnetograms with a
cadence of 30 seconds. They are constructed once every 96 minutes. The error
in the flux densities per pixel in the averaged magnetograms is
$\sim 9$~G, and each pixel has a mean area of $1.96$~Mm$^2$.

  At the chromospheric level we have used full-disk H$\alpha$ data from Big
Bear Solar Observatory (BBSO) and the Optical Solar Patrol Network (OSPAN)
at the National Solar Observatory in Sacramento Peak.

   To identify changes in the low corona associated to
the source regions of the identified MCs, the Extreme-Ultraviolet Imaging
Telescope \citep[EIT,][]{Delaboudiniere95} on board SOHO is used. EIT images
chromospheric and coronal material through four filters. In particular, we
have analyzed the 195{~\AA} band which images plasma at 1.5$\times$10$^6$~K.
When available we
have used observations taken by the Transition Region
and Coronal Explorer \citep[TRACE,][]{Handy99} in the 171{~\AA} band.

   The identification of CMEs is done using LASCO on board SOHO, plus 
proxies for eruptions in the chromosphere and lower corona.
For the analyzed time interval, LASCO imaged the solar corona from
$\sim$ 2 - 30 solar radii with two different coronagraphs:
LASCO C2 (2 -- 6 solar radii) and C3 (4 -- 30 solar radii).

\subsection{Solar activity from May 11 to May 12, 2005}
\label{sol-active}

 Without any doubt, the main contribution to the
larger cloud (MC2) and other IP signatures discussed in previous sections
is the halo CME appearing in LASCO C2 at 17:22~UT on May 13
(see Sections~\ref{sol-low-corona} and~\ref{sol-upper-corona}).
However, since we have found that the ACE magnetic field and plasma 
observations
from 05:42~UT on May 15 ('3') to 10:30~UT on May 17 ('10') can be
interpreted as being comprised by two different structures
(see Sections~\ref{mcs-models-23}-~\ref{mcs-models-47}),
we describe the solar activity observed by EIT and LASCO preceding the major 
CME on May 13.

   From May 11 until May 13, 2005, solar activity
is mainly concentrated in two ARs, AR~10758 located in the southern hemisphere
and AR~10759 in the northern hemisphere, which produces the most intense 
events.
Activity progressively increases in the later region from May 11 to 12; then,
flares reach level 2B in H$\alpha$ and M1.6 in soft X-rays.

   On May 11 a CME is first seen in LASCO C2 at 20:13~UT above
the SW limb. This event is classified as a full halo.
The linear and second order fittings
to LASCO C2 observations give plane-of-sky (POS) speeds of
550~km~s$^{-1}$ and 495~km~s$^{-1}$, respectively (from LASCO CME
Catalog, http://cdaw.gsfc.nasa.gov./CME\_list/index.html). An
M1.1 flare in NOAA AR~10758, located at S11W51, starting at 19:22~UT can be
associated with this event. 
Subframes from the EIT shutterless campaign
show a raising feature in absorption
by 19:22~UT. However, the closeness in time/space between the
two MCs discussed in Section~\ref{insitu-icme-obs} makes it highly improbable
for this event to be the source of MC1, since by simple assumption of
constant speed it would take $\sim$ 3.1 days to reach Earth and
then would be expected to arrive before May~15, i.e. earlier than MC1.

   On May 12, an event (flare, dimming, and loops in expansion)
is seen in EIT at about 1:57~UT. This event originates in AR~10759 and is
accompanied with a C9.4 flare located at N11E31 with peak at 1:10~UT.
No CME is observed in LASCO. However, a diffuse and gusty outward flow
is evident in LASCO C2.
There is another apparent event in EIT at
about 13:56~UT, to the south of AR~10760. The related C3.0 flare occurs in
S11W62 at 13:40~UT. This event shows an association with a CME, weak and
slowly traveling in the SW direction. This CME
has a strong component in the POS and, therefore, we do not expect it to be
significantly directed towards Earth. Furthermore, from the orientation of AR~10760
polarities during their disk transit, we can infer
from MDI data that its magnetic helicity sign is positive,
contrary to that of both studied MCs
\citep[see the description and interpretation of magnetic tongues in][]
{LopezFuentes00, LopezFuentes03}.

\begin{figure*}
\centering
\includegraphics[width=0.4\linewidth]{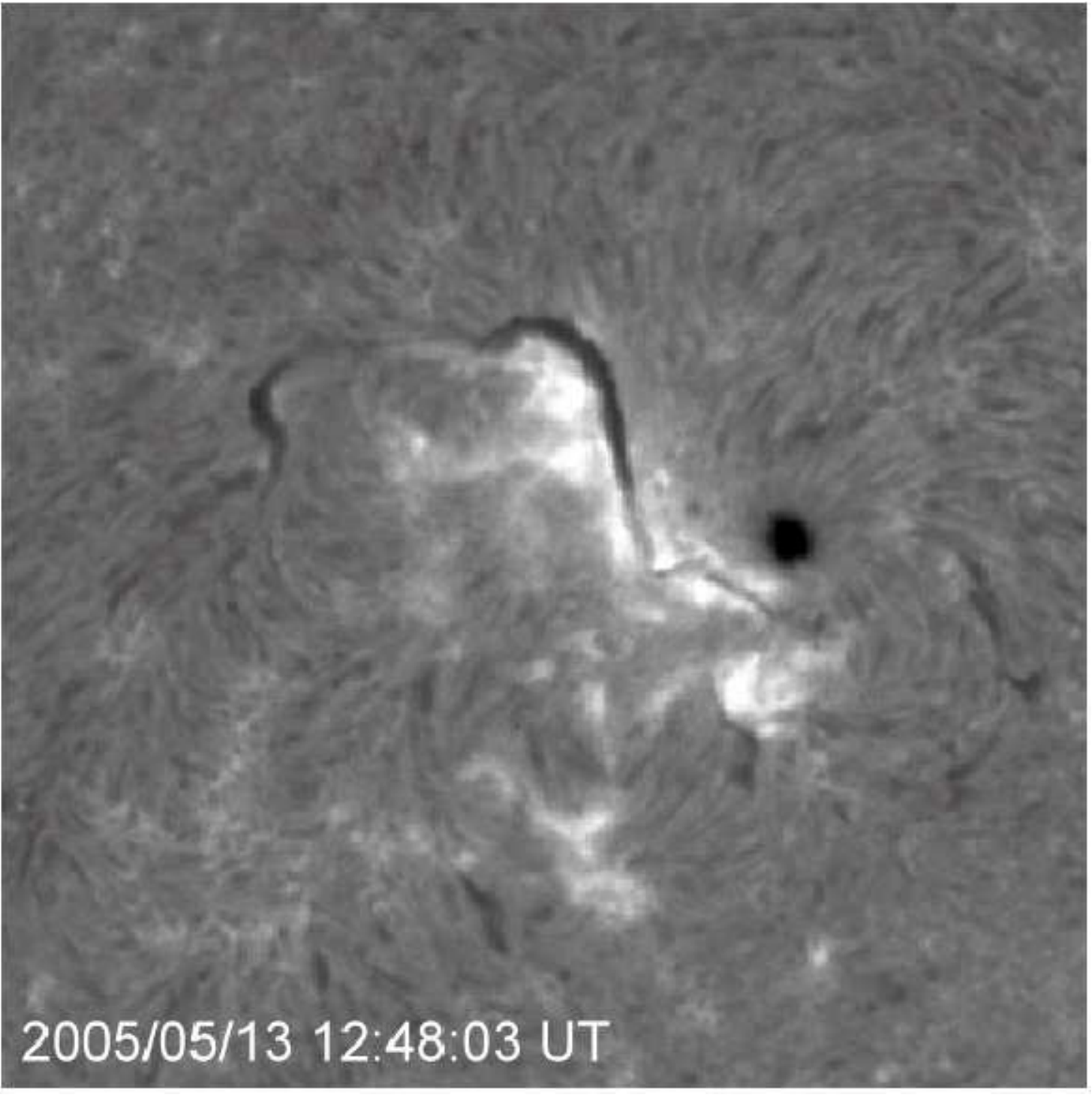}
\vspace*{0.2cm}
\includegraphics[width=0.4\linewidth]{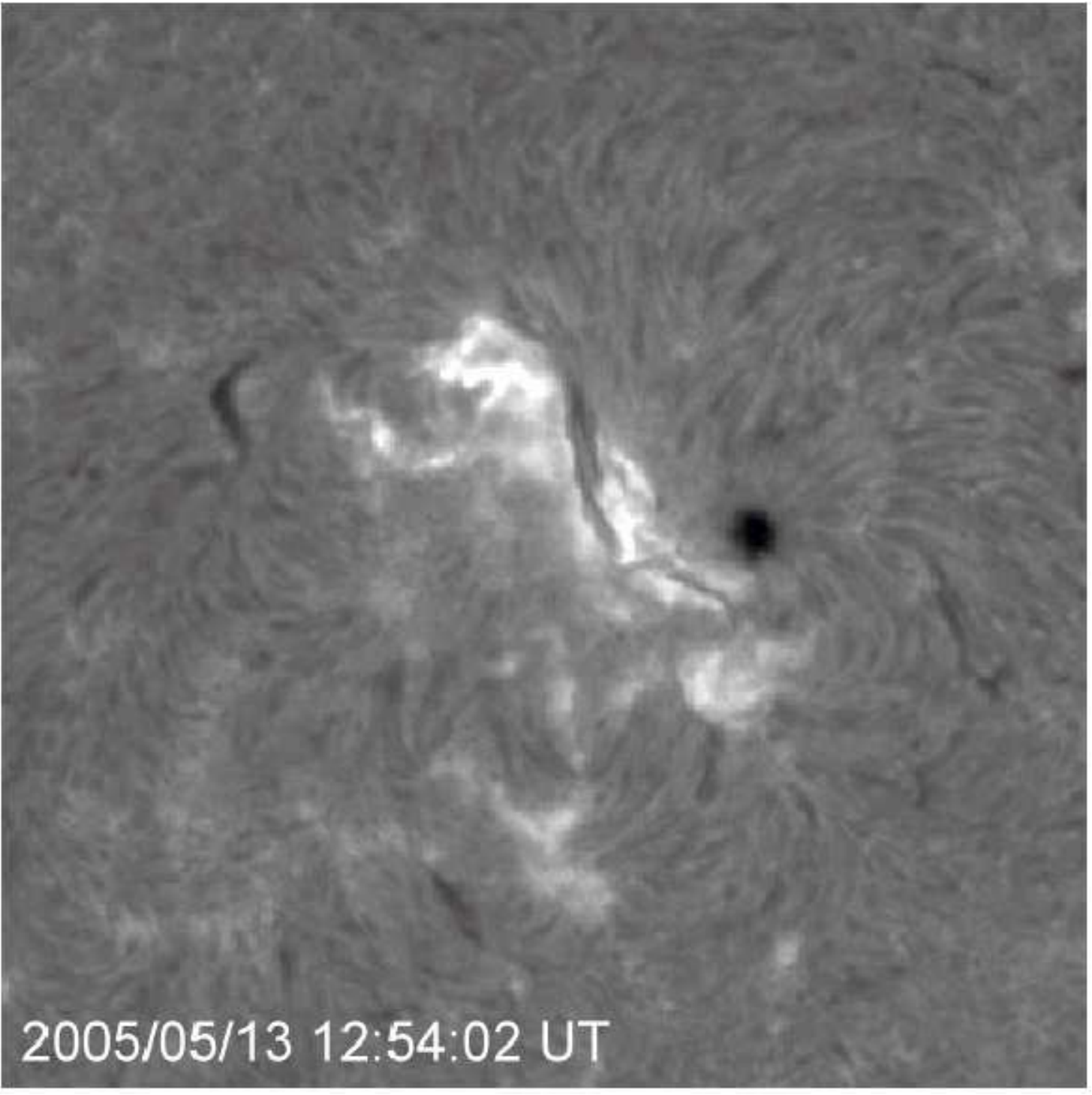}
\caption{OSPAN H$\alpha$ images of the flare at 12:49~UT. The left panel
shows the portion of the filament lying along the E-W northern
magnetic inversion line expanding northward. Flare ribbons and flare
loops
linking the ribbons are visible in the right panel.
The E-W middle portion of the filament is no longer visible.}
 \label{fig-halpha}
\end{figure*}

   After discarding these previous events, and considering the closeness in
time/space between the two MCs analyzed in Section~\ref{solar-wind}, we
concentrate in the activity observed during May 13.
In the next sections we will present observations and discuss, in particular,
two flares and associated filament eruptions
that we consider to be the solar source events of the clouds. The flare
timings are given by GOES soft X-ray data (Figure~\ref{fig-goes}).

\subsection{May 13 events: Photospheric and chromospheric signatures}
\label{sol-chromosphere}

\begin{figure*}
\centering
{\includegraphics[width=0.32\linewidth]{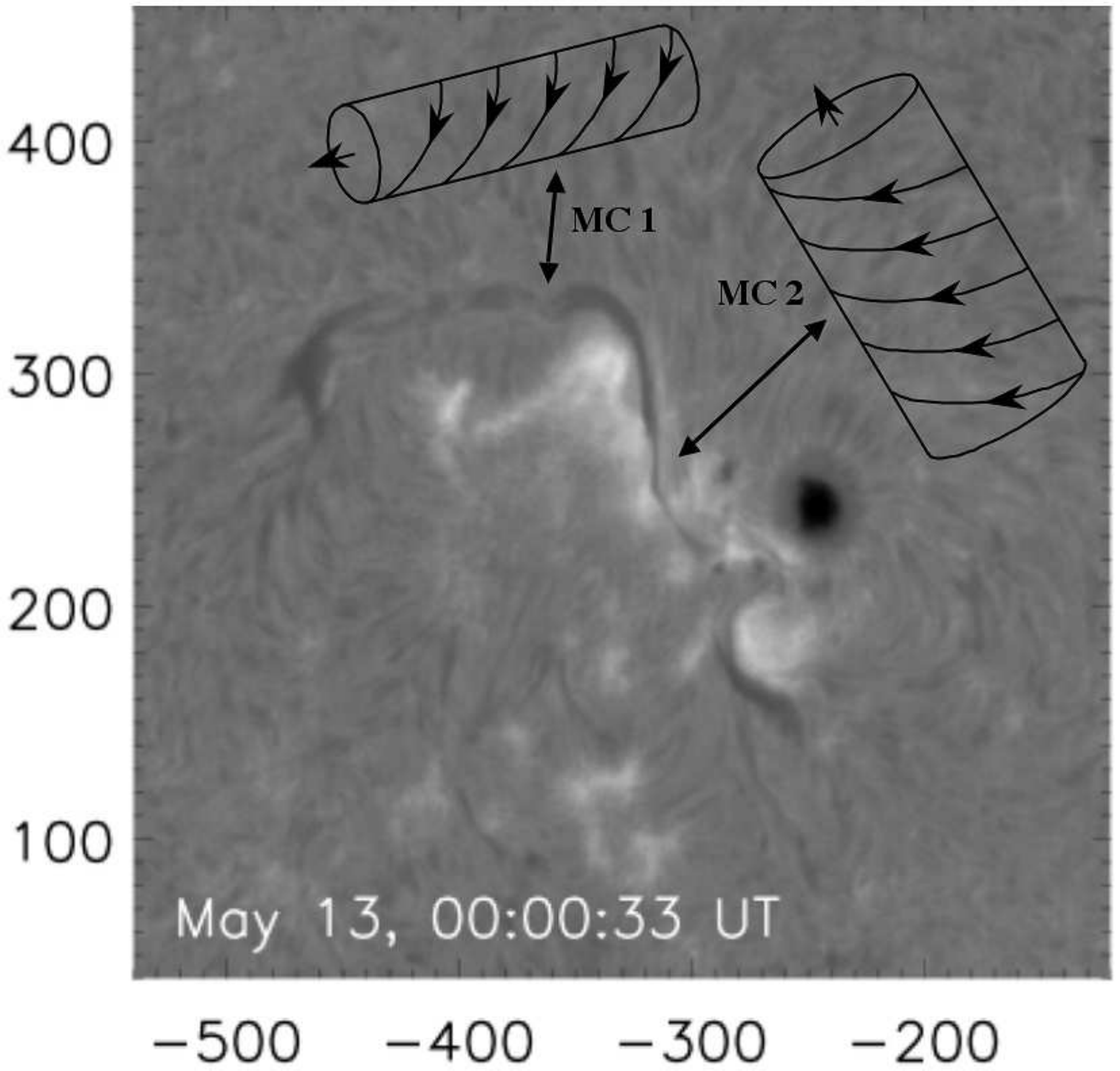}
\includegraphics[width=0.32\linewidth]{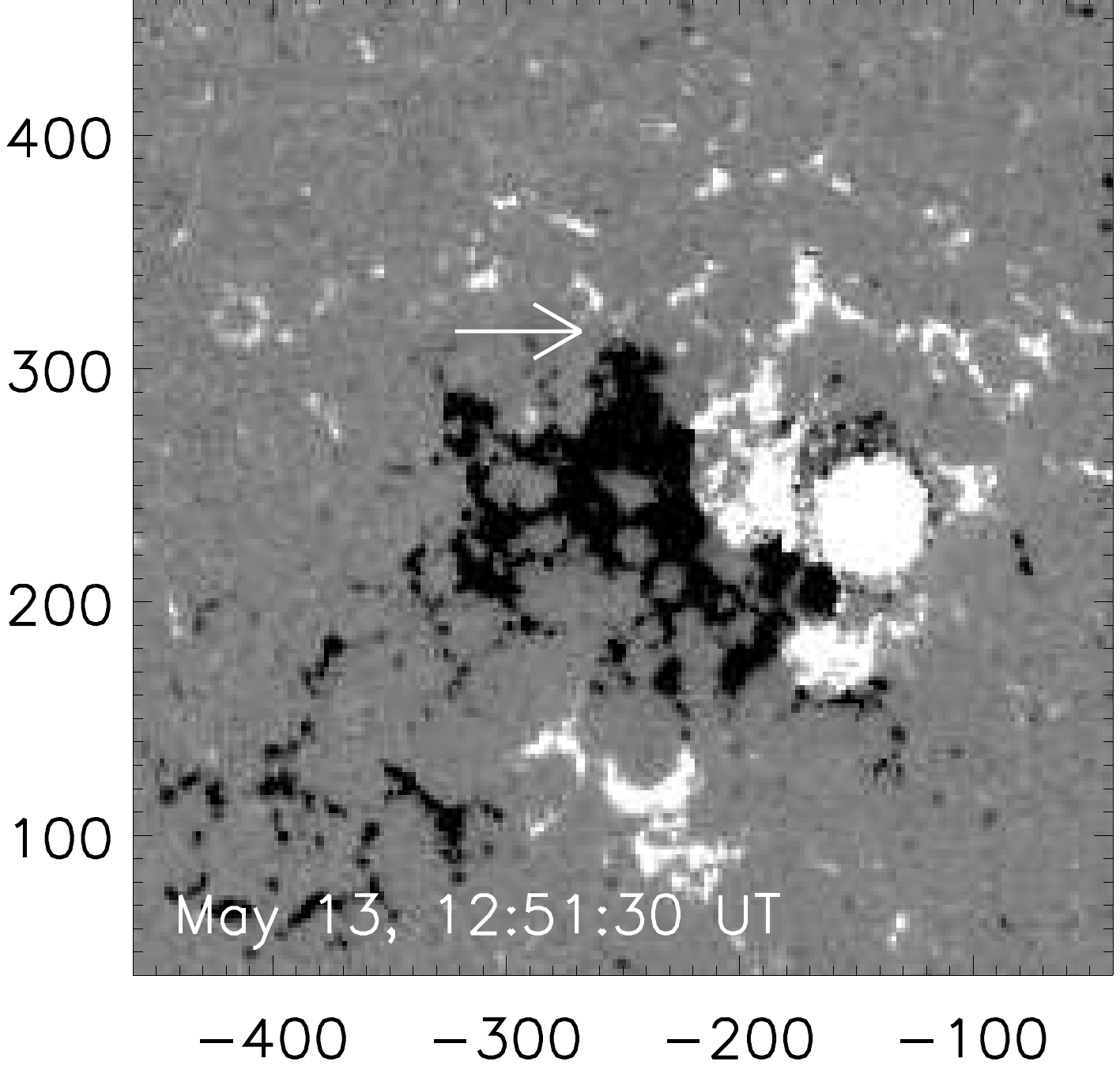}
\includegraphics[width=0.32\linewidth]{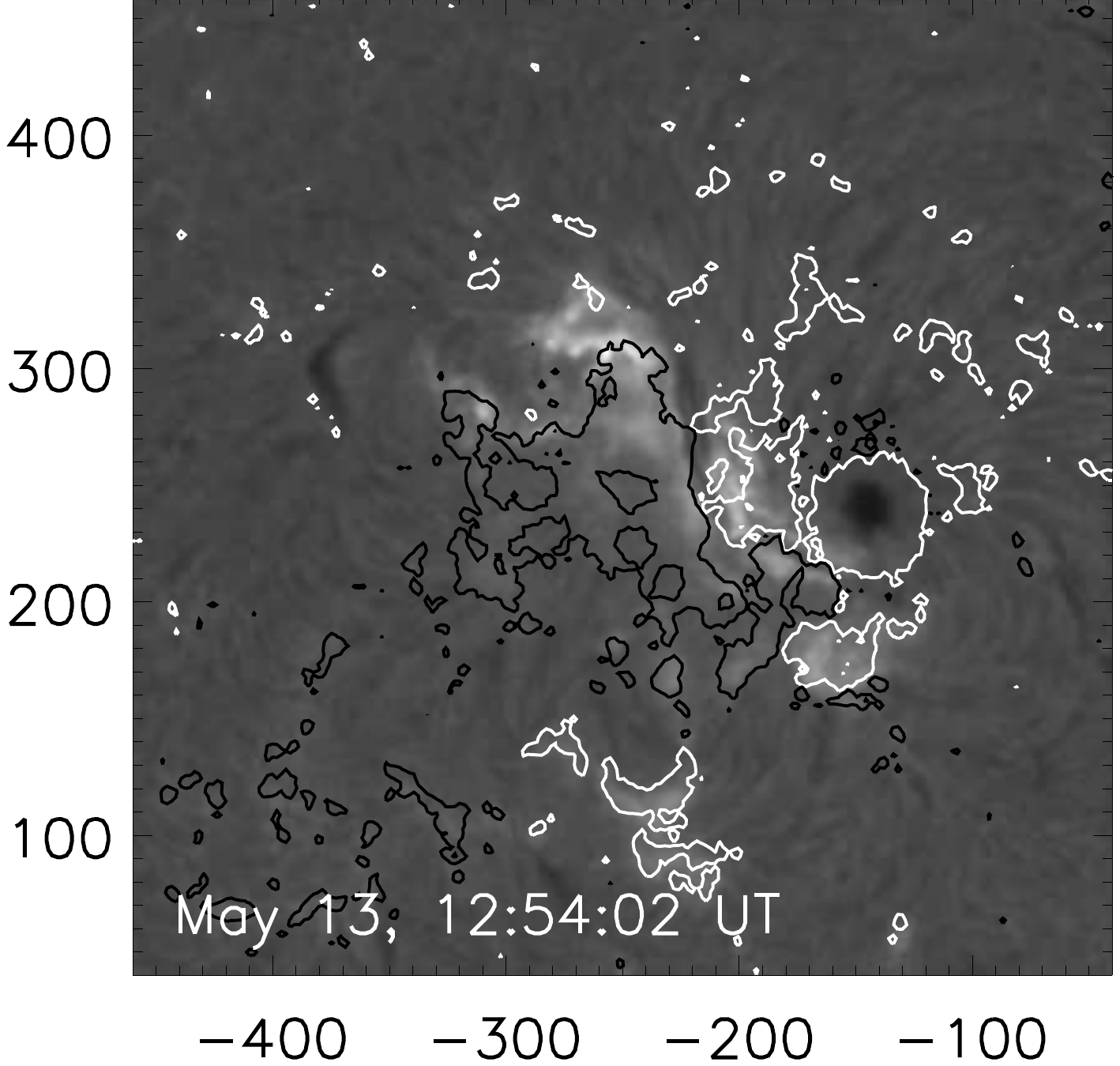}}
\caption{Evidence of flux cancellation at the northern E-W portion
of the filament channel in AR~10759.
Left: BBSO H$\alpha$ image early on May 13 showing the non-uniform shape of the
E-W filament portion. The two thick white arrows indicate
the axis orientations of the two identified MCs projected on the solar disk.
Center: MDI magnetogram closest in time to the flare at 12:49~UT
(the field has been
saturated above/below $\pm$ 100~G). The arrow points
to a small positive flux concentration that moves towards the negative
field across the filament channel. Several of such flux cancellation episodes
occur during May 13.
Right: An overlay of a flare image and the $\pm$ 50~G MDI isocontour (white
positive field, black negative field). In all images the axes are in arc sec.}
\label{fig-mag-cancel}
\end{figure*}

  Along May 13 the filament in AR~10759 is seen to activate several times,
mainly in its northern fraction. At 12:49~UT
a C1.5 flare starts in N17E15, with peak at 13:04~UT. The rising filament is
seen in H$\alpha$ and also in TRACE (see the left Figure~\ref{fig-halpha} and
Section~\ref{sol-low-corona}). At the chromospheric level a two-ribbon flare,
classified as a subflare faint, develops along the magnetic
inversion line at the north of AR~10759 (Figure~\ref{fig-halpha}). This is
the only event showing eruptive characteristics at that time on the solar disk.
This portion of the inversion line is oriented in the E-W direction.
Part of the filament extending along this E-W inversion line
is no longer visible by 12:54:02~UT.
Flare loops linking the two ribbons are also visible in H$\alpha$ at this time.
What may be the cause of the destabilization of the E-W portion of the filament?
We have analyzed the magnetic changes observed in the filament channel
previous to the flare at 12:49~UT, for this we use MDI line of sight magnetic
maps. Continuous flux cancellation is observed at the magnetic inversion line
along
which the erupting part of the filament lies.
The E-W northern fraction of the filament shows a non-uniform shape
(as being formed by several sections) and some barbs or feet are present
(Figure~\ref{fig-mag-cancel} left). A small positive polarity is seen to
intrude
in the negative field in between two sections of the filament (see the arrow in
Figure~\ref{fig-mag-cancel} center); this single intrusion implies a
cancellation of $\sim 5\times 10^{19}$ Mx. The close relationship between the
flare and the
magnetic cancellation site is shown in the right panel of
Figure~\ref{fig-mag-cancel}. MDI low temporal and spatial resolution does not
allow us to
follow in detail the evolution of the small flux concentrations and ascertain
the clear association between their cancellation and filament eruption;
however, we believe this to be the most plausible cause.
Small-scale magnetic changes in flux concentrations along filament channels
have been frequently reported a few hours prior to local filament restructuring
\citep[see e.g.,][]{Chae00,Deng02,Wood03,Schmieder06b}.
Furthermore, \cite{Zhang01} attributed
the origin of the major solar flare and filament eruption on July 14,
2000, to flux cancellation at many sites in the vicinity of the active
region filament. 

  Taking into account that the C1.5 flare occurs by only about 4 hours earlier
than the main flare and eruption in AR~10759 (see below) and
the similar orientation of the
MC1 axis and of the magnetic inversion line along which the erupting
filament lies (Figure~\ref{fig-mag-cancel} left), we consider that the
partial eruption of the AR filament is the source of this first cloud.
There is evidence for the breakup of filaments into more than one segment,
\cite[see e.g.][]{Martin72}
and a recent example in \cite{Maltagliati06}.
Even some filaments erupt only partially, or saying it in a different way different portions of
the same filament may erupt at different times and trigger different flares
\citep{Martin72,Tang86,Maltagliati06,Gibson06a,Gibson06b,Liu08b}.
This kind of eruptions have been discussed in the frame of the eruptive
flux rope model for CMEs \citep[see the review by][]{Gibsonetal06}.
Moreover, as discussed below in
Section~\ref{sol-upper-corona}, on May 13 there are no other evidences of
CMEs in LASCO until 17:22~UT.

  The solar event following the one just described is the major M8.0/2B flare
that starts at 16:13~UT and peaks at 16:57~UT. This flare ended on May 14 at
$\sim$ 17:00~UT, being a long duration event (see Figure~\ref{fig-goes}). The
photospheric and chromospheric observations corresponding to this event
have been shown and discussed by \cite{Yurchyshyn06} and \cite{Liu07}.
The former authors have associated
this flare and eruption to a single MC comprising the two we have
identified in this paper. We want to stress here that the direction of MC2 axis
is in agreement with the
orientation of the N-S fraction of the AR filament that lies along the main
magnetic inversion line (see Figure~\ref{fig-mag-cancel}, left panel).

\begin{figure*}
\centering
\includegraphics[width=0.23\linewidth]{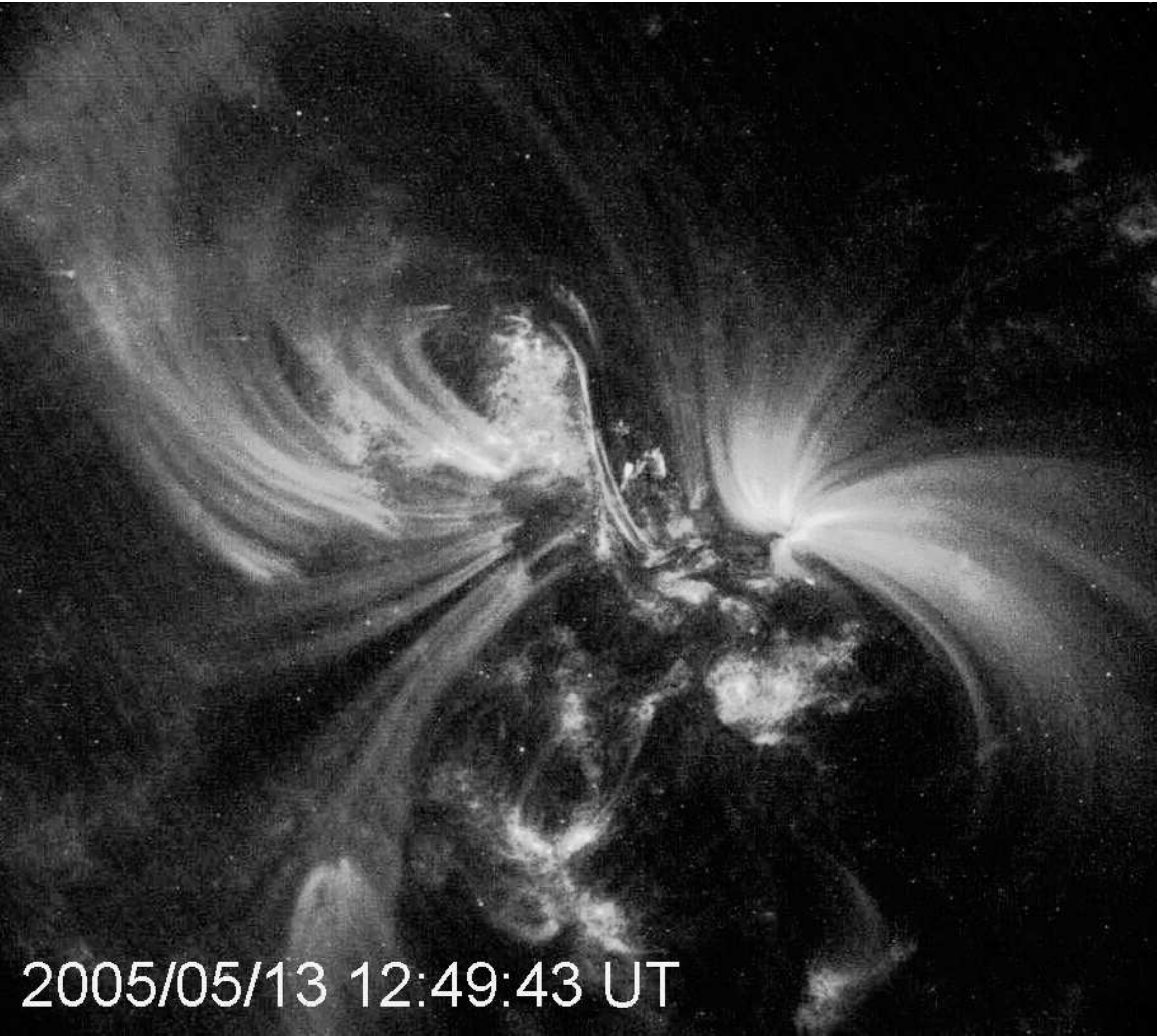}
\vspace*{0.2cm}
\noindent\includegraphics[width=0.23\linewidth]{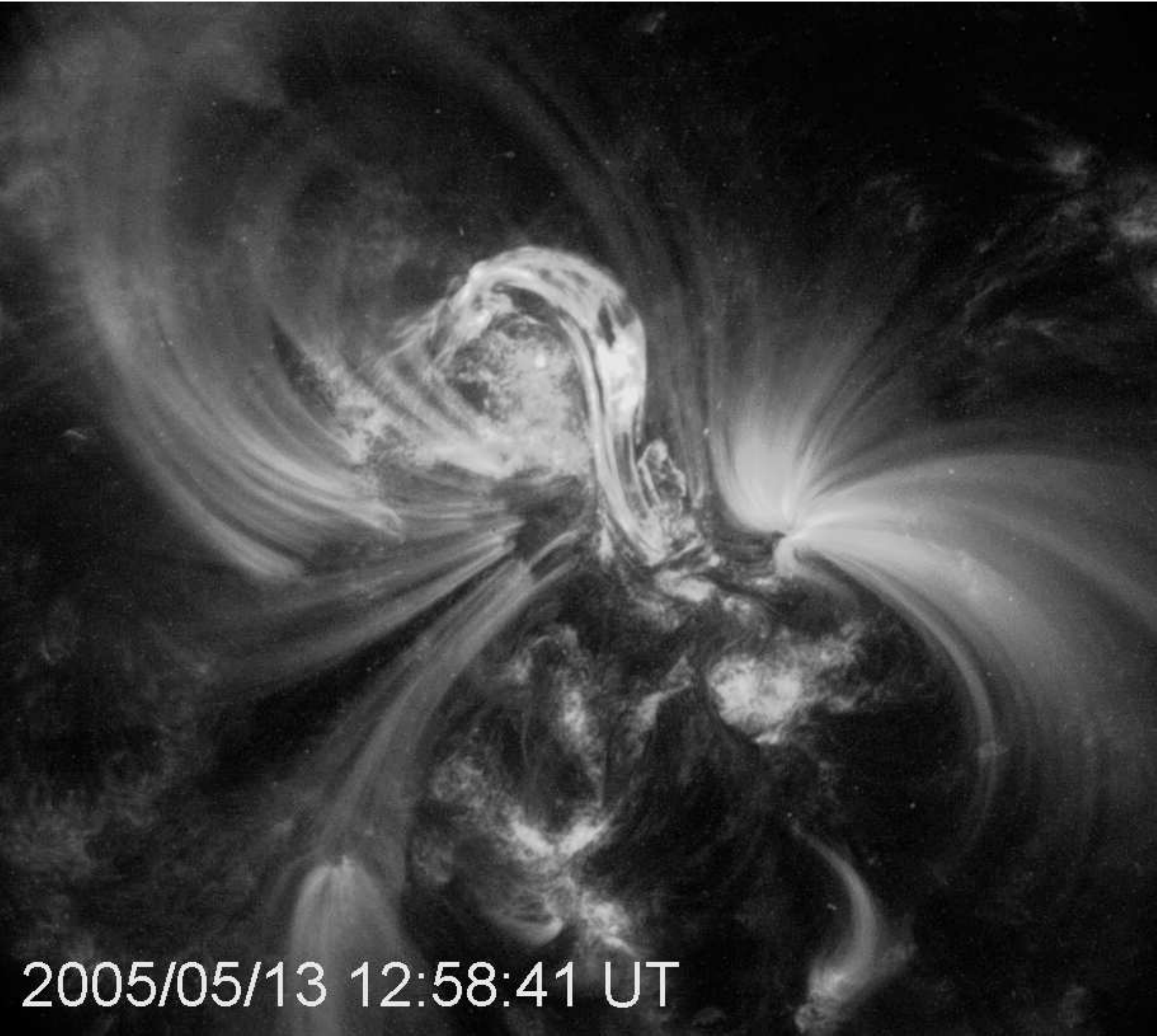}
\vspace*{0.2cm}
\noindent\includegraphics[width=0.23\linewidth]{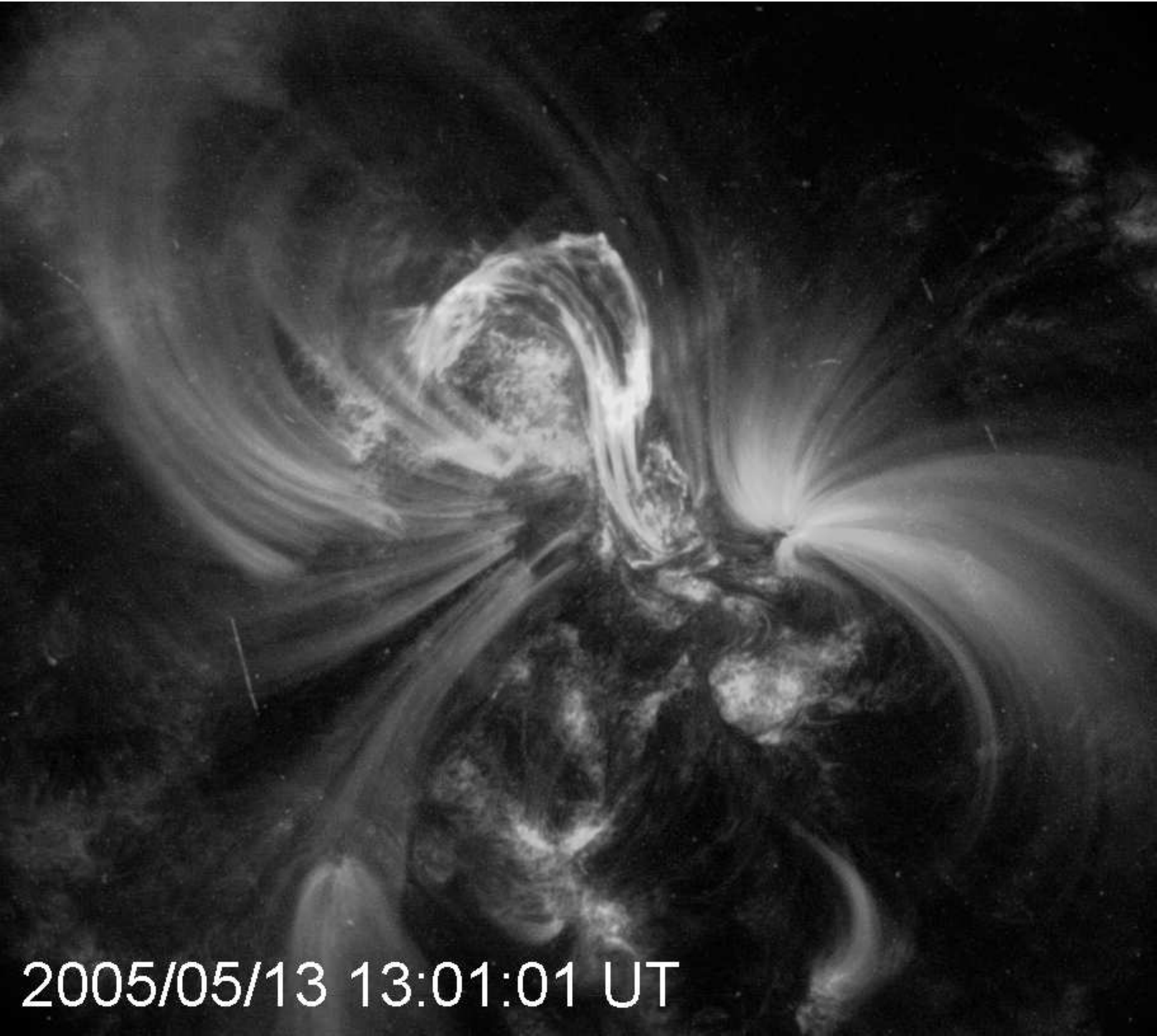}
\vspace*{0.2cm}
\noindent\includegraphics[width=0.23\linewidth]{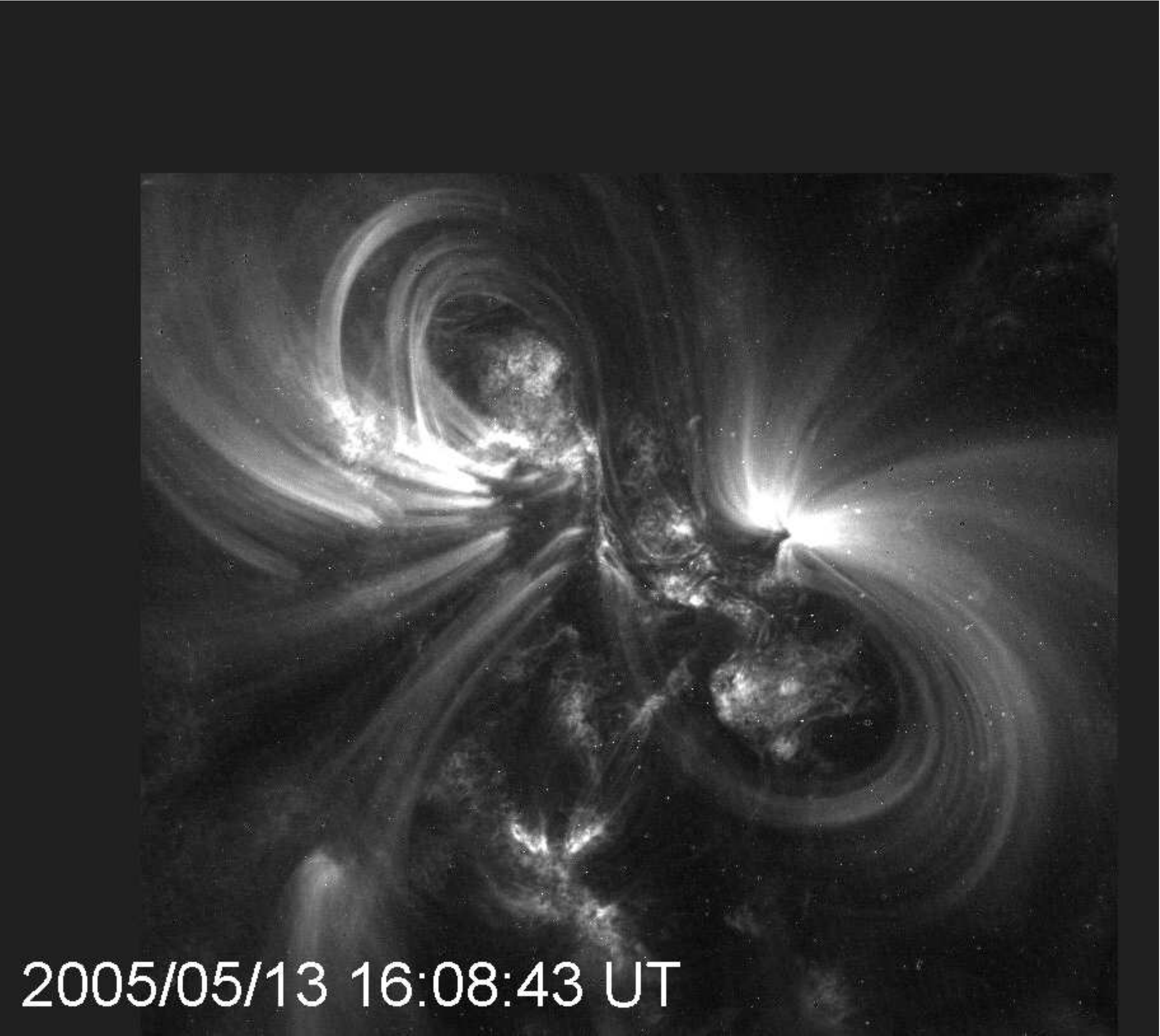}
\caption{TRACE images in 171{~\AA}.
The first three panels show different stages of the
expansion of the filament embedded in a brighter curved structure during
the flare at 12:49~UT. The last panel shows that the
coronal loop structure recovers its global shape after the C1.5 flare and
before the M8.0 flare at 16:32~UT.
The same region is showed in the four panels, though the TRACE field of
view is shifted for the image at 16:08:43~UT.}
\label{fig-trace}
\end{figure*}

\subsection{May 13 events: The lower corona signatures}
\label{sol-low-corona}

\begin{figure*}
\includegraphics[width=0.5\linewidth]{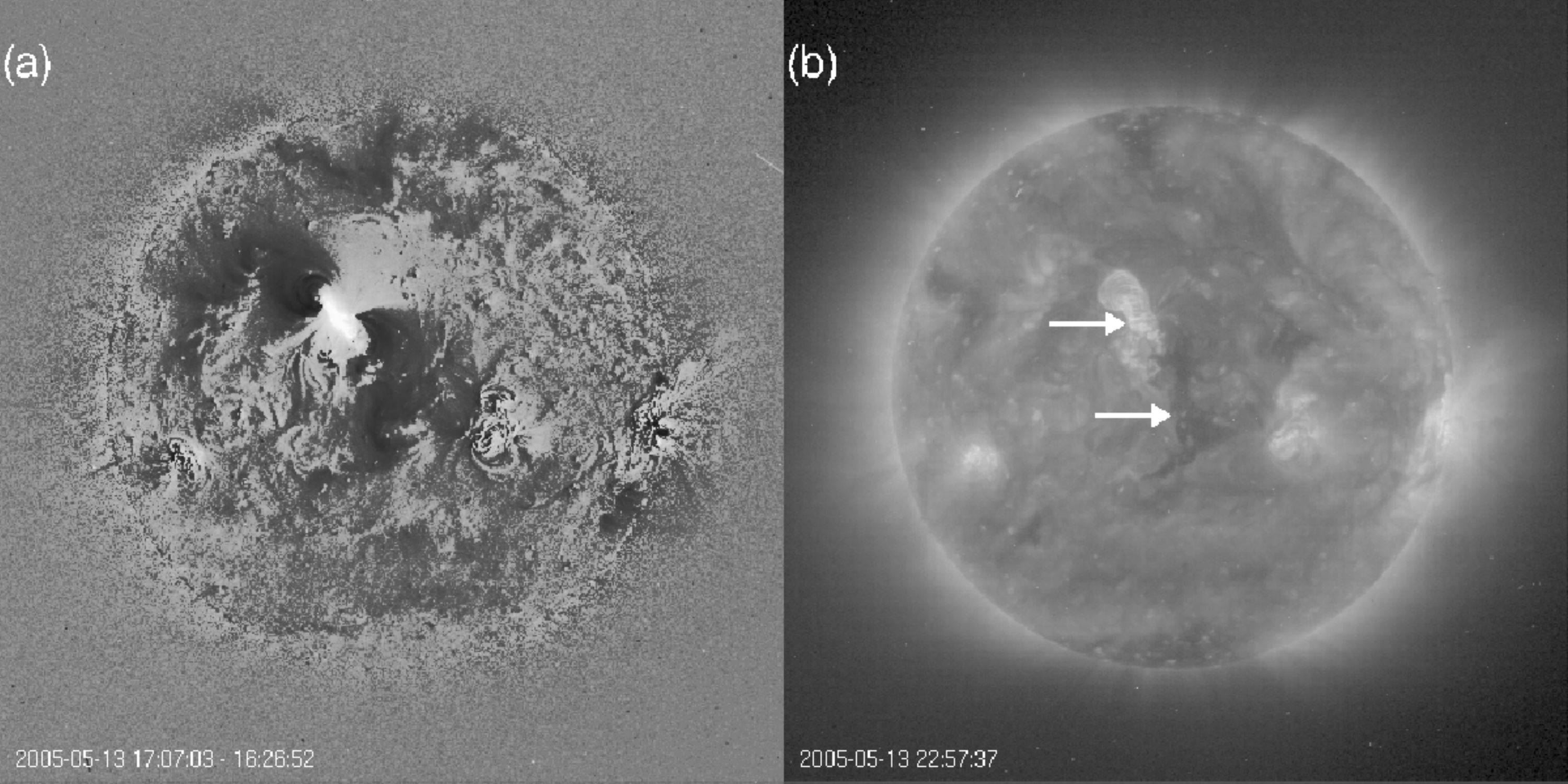}
\caption{(a) SOHO/EIT difference image in the Fe~XII bandpass (195~\AA)
showing a large coronal dimming (dark areas) corresponding to the CME
associated with the M8.0 flare on May 13, 2005. The last pre-eruption
image
taken at 16:26~UT was subtracted from the image taken at 17:07~UT. (b)
SOHO/EIT image in the Fe~XII bandpass (195~\AA) taken at 22:57~UT on May 13,
2005 (after the start of the CME associated with the M8.0 flare).
Post-eruption arcade and a transient coronal hole are marked with top and
bottom arrows respectively.}
\label{fig-dimm}
\end{figure*}

  In the lower corona, TRACE observes the event at 12:49~UT in 171{~\AA}
(Figure~\ref{fig-trace}).  Images before 12:39:49~UT and after 13:04:27~UT are
not useful because of a strong particle shower.
TRACE data are dominated by a large-scale sigmoidal structure
(with an inverse S shape) previous to the flare well visible by
$\sim$ 12:40~UT.
The small scale loops linking the two H$\alpha$ ribbons
are not distinguishable in TRACE images.
By 12:49:43~UT the filament can be observed as a dark curved structure
surrounded by a brighter one, probably formed by heated filament
material (see Figure~\ref{fig-trace}). This bright structure is
seen expanding upward until 13:04:27~UT, though it is hard to assure
that it erupts because of lack of visibility in the following images (Figure~\ref{fig-trace},
central panels).

  Later on, by $\sim$ 16:00~UT (last panel in Figure~\ref{fig-trace})
and before the M8.0 flare, TRACE
images appear again dominated by the large-scale sigmoid.
This structure is similar to the one described by \cite{Liu07}
seen before the M8.0 flare; in their Figure 1 they show the different
parts that constitute the sigmoid: the magnetic elbows, the envelope loops
\citep[according to the nomenclature of][]{Moore01} and the footpoints of the
sigmoidal fields.
Thus, the rising of the bright curved structure, shown in the central
panels of Figure~\ref{fig-trace}, is most likely linked to the filament eruption during
the first and less intense two-ribbon flare that we associate with MC1.
\cite{Yurchyshyn06} have suggested that the bright structure seen in
the central panels
is the core of the sigmoid that later starts a slow ascent (by $\sim $12:55~UT)
and become active during the M8.0 flare, but in fact it erupted earlier and
reactivated later.

  The major eruptive event, linked to an M8.0 flare, is visible in EIT
images starting from 16:37~UT (see Figure~\ref{fig-dimm}).
It exhibits several CME-associated phenomena \citep{Hudson01}:
coronal dimmings, a large-scale EIT wave and a post-eruption arcade.
As it was pointed out by \cite{Liu07}, the darkest part of the dimming
comprises two patches to the NE and SW of the erupting active region.
The SW dimming is associated with a transient coronal hole
(shown by an arrow in Figure~\ref{fig-dimm}b).
It was suggested that transient coronal holes can be
interpreted as footpoints of the ejected flux rope
\citep[see, e.g.,][]{Webb01,Mandrini05}.
The absence of the clearly visible second transient
coronal hole in this event may indicate that the second leg of the erupted
flux rope may have already reconnected with the ambient coronal magnetic field
\citep[this reconnection changes the very extended field lines rooted in the dimming by closed magnetic connections][]{Attrill06,Kahler01}.
The overall morphology of twin dimmings and the
presence of the post-eruption arcade (Figure~\ref{fig-dimm}b) suggest that an
eruption of a coronal magnetic flux rope had taken place
\citep[e.g.,][]{Hudson01}.

\subsection{May 13 events: The upper corona observations}
\label{sol-upper-corona}

\begin{figure*}
\centering
\includegraphics[width=0.7\linewidth]{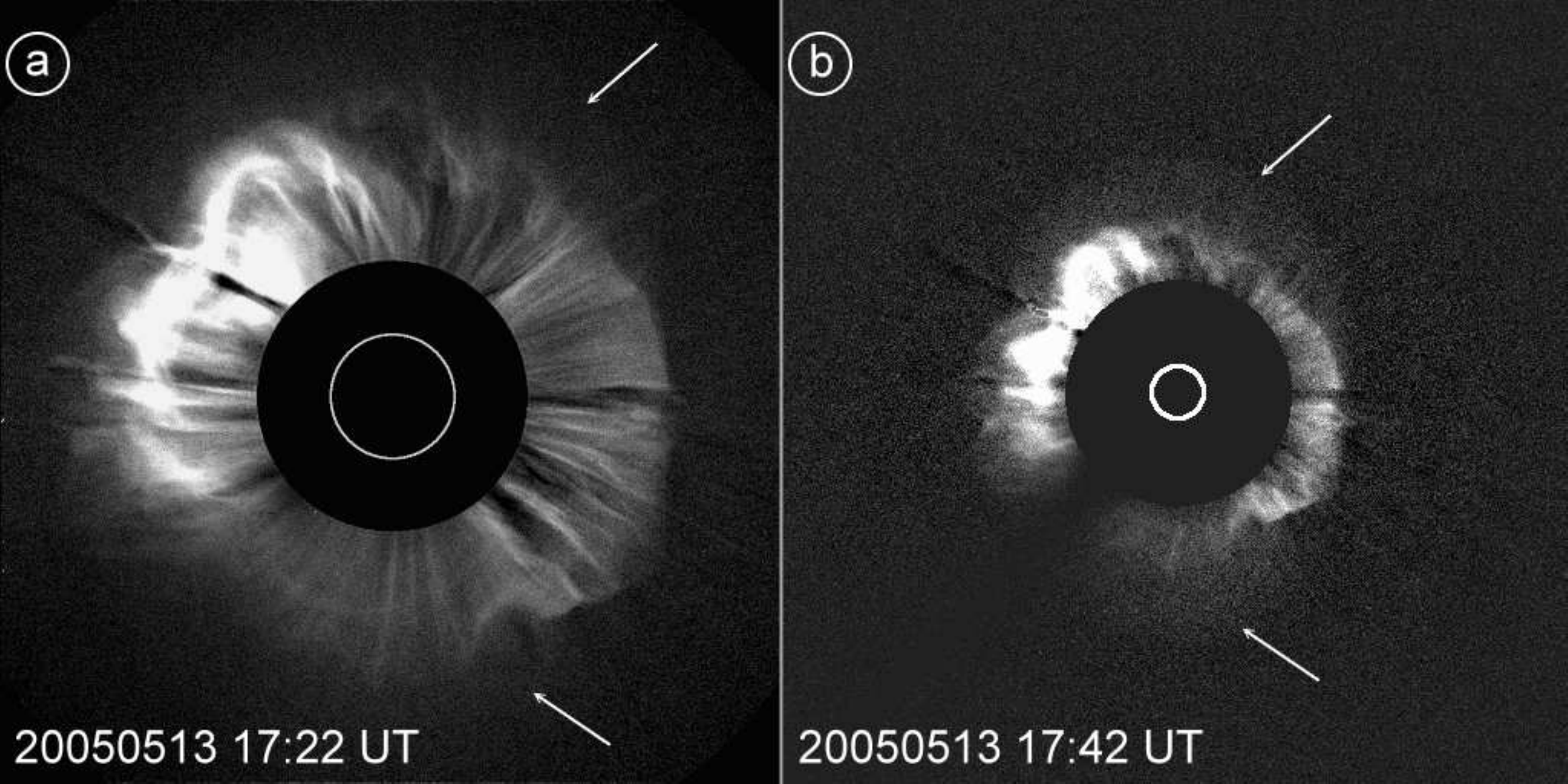}
    \caption{LASCO images of the halo CME on May 13, 2005, as recorded by
a) LASCO C2 and b) LASCO C3. A pre-event image has been subtracted in
both cases. These are the two only full-field of view images taken of
the event. The arrows indicate the edge of the faintest front.}
\label{fig-lasco}
\end{figure*}

  On May 13, 2005, a full halo CME with brightness asymmetry towards the NE
first appears in the field of view of LASCO C2 at 17:22~UT
(see Figure~\ref{fig-lasco}a).
Five more partial snapshots of the halo CME were captured by LASCO C3 from
17:20 to 17:50~UT. The closeness in time of the latter images does not allow
to see major changes from one to another, so we present here only one of them
(Figure~\ref{fig-lasco}b).
The estimated POS speed for the fastest front, as calculated by
the LASCO CME catalog, is $\sim$1690~km~s$^{-1}$. Further information on
the kinematics of this CME and its associated shock can be found in
Section~\ref{radio-obs} and in \cite{Reiner07}.

  Enhancement of the LASCO images presented in Figure~\ref{fig-lasco}
allows to distinguish a very faint edge toward the south and north, about a
couple of solar radii ahead of the brightest one (indicated by arrows).
This feature suggests the presence of a shock, as described
by \cite{Vourlidas03} and \cite{Vourlidas06}.
Finally, the CME related to the solar eruption at 12:49~UT is probably so
faint, that its detection was not likely due to sensitivity limitations
of LASCO
\citep[e.g., ][]{Tripathi-etal2004,Yashiro05}.
Furthermore, \cite{Yashiro05} conclude that all fast and wide CMEs are
detectable by LASCO, but slow and narrow CMEs may not be visible when they originate
from disk center.
In addition to this restriction, which impedes the visualization of a CME
if it is feeble, the LASCO cadence during and after the C1.5 flare at 12:49 UT
was of 30 minutes, as opposite to the usual 12 minutes of LASCO C2.

\section{Two structures in the inner heliosphere: Remote radio type~II
observations}
\label{radio-obs}

\begin{figure*} %
\centering
\includegraphics[width=\linewidth]{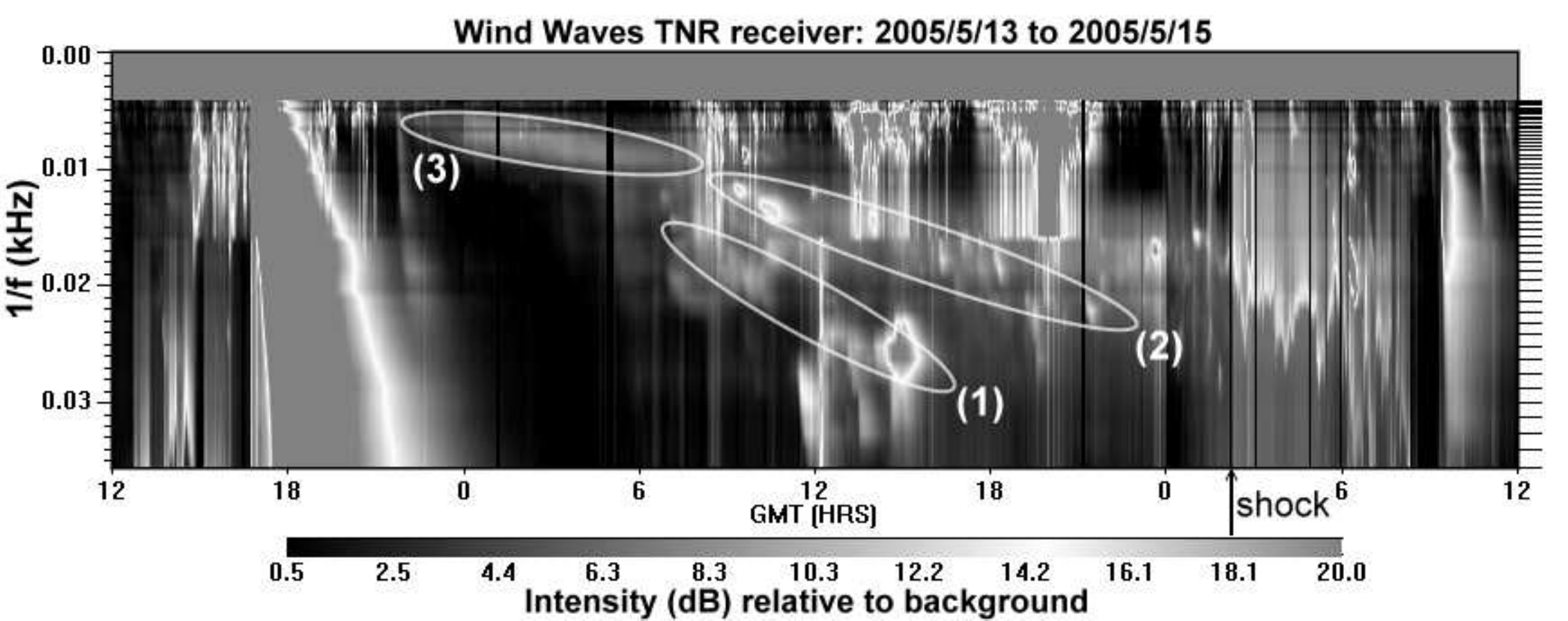}
    \caption{Dynamic spectral plot showing the radio emission recorded
    by the WAVES/TNR receiver on board Wind between May 13 at 12:00~UT
    and May 15 at 12:00~UT.  The vertical axis represents 1/$f$ [kHz$^{-1}$]
    while the horizontal axis corresponds to time.
    Three identified type~II emissions
    are labeled (the entity (a) is associated to (1) and (2) which are the
    fundamental and harmonic emission at the plasma frequency, and the entity
    (b) is associated to (3)). The black arrow in the horizontal axis denotes
    the shock arrival time. Tick marks on the right vertical axis indicate the
    location of TNR frequency channels for the plotted frequency range.
}
    \label{fig-radio}
\end{figure*}

When a halo CME leaves the field of view of a coronagraph, it is occasionally
possible to further track its evolution by means of measurements in the
radio regime of the electromagnetic spectrum. The shock commonly driven
by a CME excites electromagnetic waves emitted at the
local plasma frequency ($f \sim \sqrt{n_{e}}$) and/or its harmonic. As the shock
travels outward from the solar corona, the local plasma frequency decreases,
and a drifting signal, a type~II radio burst, is generated. This brand of
emissions
can be followed drifting in frequency from $\sim150$~MHz down to $\sim25$~kHz,
which is the approximate value of the local plasma frequency in the vicinity
of Earth \citep{Leblanc98}.
Then, the shock traveling toward Earth is detected by instruments
in the IP space, such as WAVES on board Wind \citep{Bougeret95}.
The ICME can also be detected by ground-based antennas measuring interplanetary
scintillation of radio sources \citep[e.g.,][]{Manoharan06}.

  The metric components of type~II bursts have long been used by space weather
forecasters as input to models of shock arrival time prediction
\citep[e.g.,][]{Dryer84, Smith95, Fry01}. Nevertheless, there is ongoing
debate on the mechanism that generates them, questioning their direct
relationship with
shocks ahead of CMEs \citep[e.g.,][]{Pick99, Gopalswamy01, Classen02, Cane05}.
Therefore, \cite{Cremades07} attempted to improve space weather forecasts
by employing information of type~II radio bursts in the kilometric
domain ($30-300$ kHz), which are indeed believed to be generated by IP
shocks \citep{Cane87}.
Accordingly, the slope of the drifting radio emission can be taken as a proxy
for the associated shock speed \citep{Reiner98}. Furthermore, for each point of
emission detected in the space-borne radio receivers, it is possible to obtain a proxy
of the radial distance from the Sun through density models
\citep[e.g.,][]{Saito77,Leblanc98,Hoang07}.
The technique used to characterize the temporally-associated radio signals makes
use of the density model developed by \cite{Leblanc98}.
To obtain the approximate sites of the emissions, and consequently a proxy of
the shock speed, it is only needed to enter into the density model an estimate
of the plasma density near Earth, usually between 5 and 9 cm$^{-3}$.
The top frequency taken into account as input for the
technique (300 kHz for the fundamental emission) corresponds approximately
to a distance of 20 solar radii according to that heliospheric density model.

We assume that at those heights most of the deceleration has already taken place,
and thus that the slope of the linear profile in Figure~\ref{fig-radio}
is enough to characterize the kinematics of a shock that travels at a constant speed.
Beyond 20 solar radii, the corresponding frequency channels of the WAVES/RAD1
detector are scarce in comparison with those of WAVES/TNR
(see rationale at the end of this Section). That is why the main goal of the
latter detector is the tracking of radio phenomena up to the orbit of Earth.
Further details on the technique, its application and success relative
to other methods can be found in
\cite{Cremades07}.

After the solar events on May 13, a number of type~II radio features
in the kilometric domain were observed by the Thermal Noise Receiver (TNR)
experiment on WAVES.
Figure~\ref{fig-radio} displays the radio intensity between May 13 at 12:00~UT
and May 15 at 12:00~UT, and from $f \sim$250 kHz to $f$=20 kHz.
In addition, each of the identified low-frequency type~II emissions has been
labeled as follows: (1) a radio emission
drifting at a speed of 1400~km~s$^{-1}$, (2) an emission drifting at a
speed of $\sim$1450~km~s$^{-1}$ and
certainly being the harmonic of the first feature, and (3) a slower radio
emission drifting
at a speed of $\sim$850~km~s$^{-1}$. The application of the technique
described in \cite{Cremades07}
to emission (1) yielded for its associated shock a predicted arrival time
at $\sim$02:40~UT on
May 15, which is in good agreement with the real shock arrival at
Earth (02:11 UT).
The arrival time derived from radio signal (2) yielded $\sim$23:20~UT on May 14. 
The error of $\sim$3 hours might be attributed to the 'patchiness' of the emission.
This effect arises because a single shock may drive type II bursts
at multiple places along the shock surface, with the possibility
of a large non-radial component of the shock velocity.
In addition, there could be significant differences in the
drift rates if the density gradient or the local shock speed varies along the
shock surface. Under these circumstances entities (1, 2) and (3) could
be the manifestation of the shock driven by only one ICME.
However, we have no data to justify any of these hypothesis.
Moreover, the large difference in propagation speed between
entities (1, 2) and (3) suggests that they must be indeed produced
by two distinct structures traveling in the solar wind or
by the same structure traveling with a significantly slower speed
at a first stage of its journey and with a faster speed at its second stage.
We will interpret this second possibility in the frame of the global observations
in next section (Section~\ref{physical_scenario_discu}).
Finally, the prediction of the arrival time of entity (3) yielded $\sim$10:20~UT on May 15.
The linear backtracing of
this last emission in time matches well with the earlier solar event on May 13.
On the other hand, entities (1) and (2) show good temporal association with the
occurrence of the second solar event on May 13 (at 16:57 UT).

Based on the observations of WAVES/TNR we present here a different
interpretation of the sources of the radio signatures than \cite{Reiner07},
who used radio data acquired by the WAVES/RAD1 detector
(see next paragraph). The TNR observations indicate the existence of 2 interplanetary
shock entities, which from now on will be referred to as entity (a) -the shock traveling
at $\sim$1400~km~s$^{-1}$,
corresponding to (1) and (2)- and entity (b) -the shock traveling at
850~km~s$^{-1}$, related to (3).
Entity (b) presumably draws near entity (a) around 07:00~UT on May 14,
at an approximate distance of 98 solar radii from the Sun. There are no
visible signs of 'cannibalism' \citep{Gopalswamy01b} in radio data, likely
due to the weakness of
entity (b) and/or because there is no significant reconnection between the
two entities.
Moreover, it is worth to note that entity (a) was not visible before 06:00~UT
on May 14,
as opposite to the slower entity (b). A plausible reason for this effect is
that entity (a) may
have been moving in an environment plasma faster than the typical solar wind,
therefore without leading shock formation.

  The kinematics analysis performed for the radio event on
May 13-15, 2005 by \cite{Reiner07} takes into account entity (a) only.
Those authors present spectral plots of the event that reach down in frequency
up to the coverage of the WAVES/RAD1 detector.
As seen in Figure~\ref{fig-radio} of that study,
entity (b) is not obvious from RAD1 measurements.
However, it does show up in TNR data, suggesting that \cite{Reiner07} did not consider
those data in their analysis. It is most likely that the poor frequency resolution
of RAD1 below 256 kHz is guilty of overseeing entity (b).
The RAD1 detector employs at a time only 32 frequency channels to monitor
the range 20-1040 kHz, using only 21 channels below 256 kHz and interpolating
to obtain the rest of the data. This may lead to false apparent broadband emissions
and to the oversight of some narrow-band features in the RAD1 domain.
Conversely, the TNR receiver employs 96 channels to cover the frequency range of 4-256 kHz,
over five logarithmically spaced frequency bands
\cite[][also see location of frequency channels as tick marks on the right vertical
axis of Figure~\ref{fig-radio}]{Bougeret95}.
This configuration achieves better frequency resolution at distances greater than $\sim$20 solar radii,
and hence makes it best suitable to study emissions beyond those distances.

\section{Conclusion}
\label{physical_scenario_discu}

We propose here a fully consistent physical scenario for the
chain of events from May 13 to 17, 2005, which
is based on a detailed analysis
of the observations presented in previous sections.

Two solar events occurred on May 13 with a difference of $\sim$ 4 hours,
both from AR~10759.
The first one, a C1.5 two-ribbon flare that was associated with the
ejection of a part of the AR
filament extending along the E-W portion of the inversion line (north of the
AR). The second and largest one, an M8.0 two-ribbon associated with the ejection
of the filament lying along the N-S portion of the inversion line and a
CME observed in LASCO C2.

\begin{figure}
\centering
\includegraphics[width=0.55\linewidth]{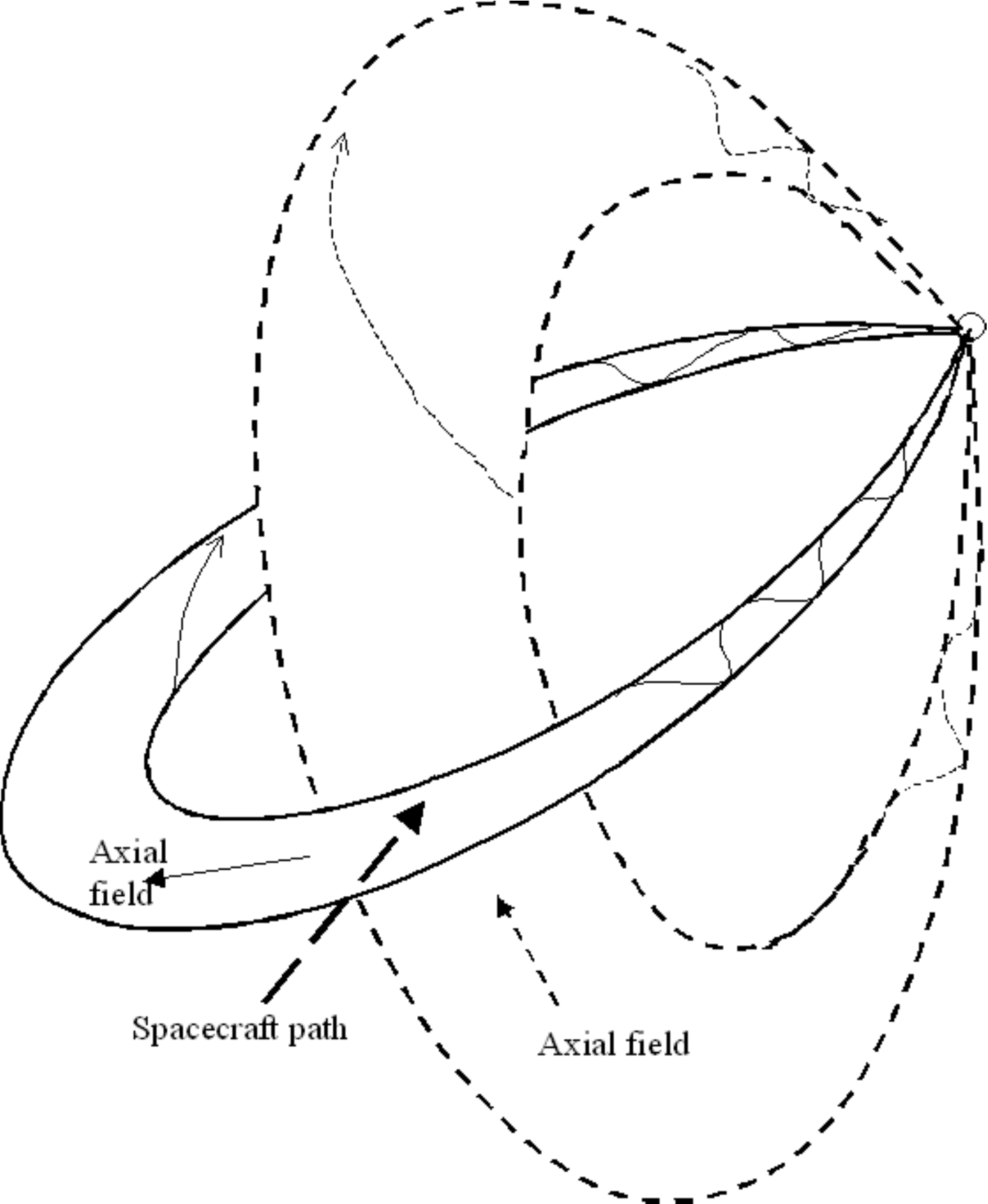}
\caption{Schematic global view of the two MCs and the trajectory
of the spacecraft that observed them.
The first MC is drawn with a continuous line (MC1),
while the second one with a dashed line (MC2).}
 \label{fig-cartoon}
\end{figure}

In the IP medium we identify an anomalously long ICME (more than 2 days).
Following the results of \citet{Burlaga01,Burlaga03}, this suggest the presence of
two flux ropes (see Sect.~\ref{insitu-icme-obs}). Indeed, we have found two clouds (MC1 and MC2).
They are stacked together and only the latitude, $\theta_{B}$, of the magnetic field permits
to identify in the data two coherent structures separated by a region (called '3-4'
in Fig.~\ref{fig-ICME-VB}) that have a different temporal evolution from neighboring regions.
Such conclusion is supported by the various attempts done to fit the data by one or two flux ropes.
Indeed, only the two flux rope models presented in Section~\ref{solar-wind}
give a reasonably good fit to the data.
Figure \ref{fig-cartoon} shows a sketch of the two clouds near Earth,
in which we have taken into account their orientations, relative sizes, and
the spacecraft trajectory during the observations.
The two MCs are oriented such that:
$\theta_{MC1}=(-12 \pm 5)^\circ$, $\varphi_{MC1}=(125 \pm 5)^\circ$,
$\theta_{MC2}=(55 \pm 10)^\circ$, and $\varphi_{MC2}=(120 \pm 10)^\circ$.
MC1 presents a low impact parameter ($p/R \sim 0.2-0.3$) and a radius
of $\sim 0.04$~AU, while MC2 is much larger and presents a larger
impact parameter; both MCs have negative magnetic helicity.
MC1 is small, has a very intense magnetic field, the signature of
an overtaking flow penetrating deep in the out bound, and almost no portion with
low proton temperature (see Figure~\ref{fig-ICME-VB}); all these observations
can be interpreted as the consequence of a compression by MC2.

The left panel in Figure~\ref{fig-mag-cancel} shows the projection on the solar
surface of the axes of MC1 and MC2 and the sense of their azimuthal field
component,
for comparison with the orientation of the erupting parts
of the filament associated with the C1.5 and M8.0 flares, respectively.
Only the second flare has an observed associated CME, and it is apriori surprising
to associate a C1.5 flare to a fast ejection (with a velocity $\approx 850$~km~s$^{-1}$
from radio data); however, it has been shown by \cite{Gopalswamy03,Gopalswamy05} that
the correlation between CME speed and flare intensity is weak.
Apart these two issues, the other characteristics of the
solar and interplanetary events are in agreement as follows.
The erupting filament orientations and directions of the field in their
associated arcades (see right panel of Figure~\ref{fig-halpha} and
Figure~\ref{fig-dimm}b)
are in good agreement with the axial and azimuthal field components in the clouds.
The AR field, as well as the two MCs have negative magnetic helicity.
Altogether, with the transit time in the expected time interval, this
demonstrates the associations between the two solar eruptions to the two interacting MCs
detected near Earth.

Moreover, an evidence of the interaction is directly found in the
interplanetary radio data, as follow.
Radio type~II observations (Section 4, entity (b)) are consistent with
a speed for
the shock driven by the first ejecta (MC1) of $\sim 850$~km~s$^{-1}$,
when it travels from the Sun up to $\sim 0.5$~AU
(from $\sim$13:00~UT on May 13 to $\sim$07:00~UT on May 14).
The shock driven by the second ejecta (MC2) seems to be weaker and
is not detected using radio data in this range of time,
probably because MC2 travels in a non-typical solar wind medium, i.e., a
medium perturbed by the previous passage of MC1, so having a faster speed.
It is also possible that this second shock
starts to interact with the trailing part of MC1 at earlier times just after
its eruption
and, thus, it turns to be weaker, as shown from numerical simulations of
interaction
of flux ropes \citep{Xiong07}.  Then, at times when MC2 reaches MC1
(at $\sim 0.5$~AU, on May 14 at $\sim$07:00~UT), when the exchange of
momentum
between the two clouds is most efficient, MC1 is accelerated and MC2 is
decelerated.
Then, both start to travel at similar speeds $\sim 1400$~km~s$^{-1}$,
but keeping their individuality.
The shock wave driven by this new combined structure is the cause
of the radio emission observed as entity (a), identified in Section 4.
During the second half of the transit to Earth, MC1 is compressed from behind 
by MC2.

\begin{figure}
\centering
\includegraphics[width=0.8\linewidth]{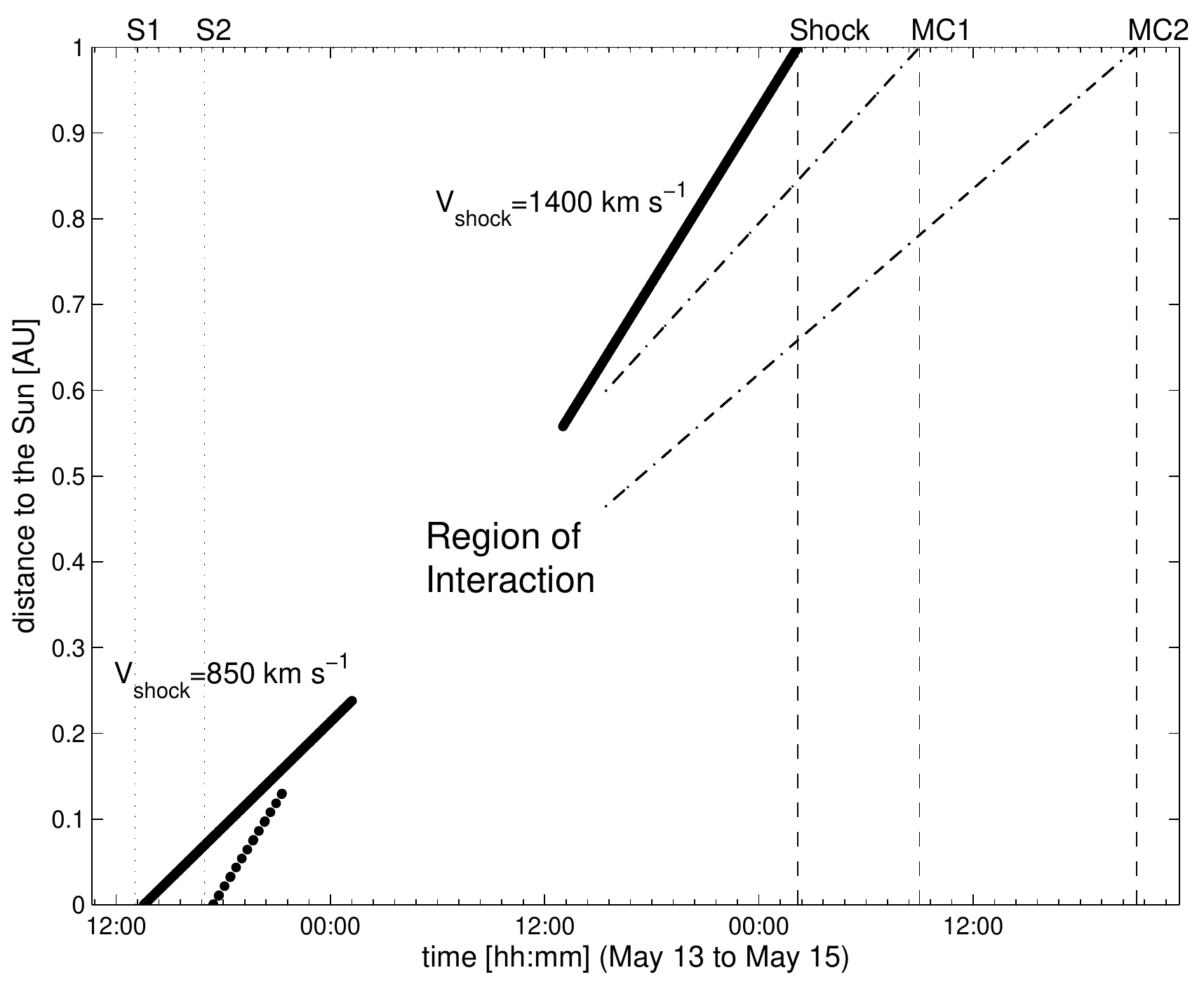}
\caption{Cartoon representing the position (as distance from the Sun along the Sun-Earth line)
of the events ejected from the Sun in function of time. Vertical dotted lines (ticks at S1 and S2)
represent the time of beginning of solar flares associated with both ejections
(May 13, $\sim$ 13:00 UT and $\sim$ 16:00 UT, respectivelly.
Vertical dashed lines (ticks at Shock, MC1, and MC2)
represent the arrival of the shock (May 15, 02:11 UT) and of the center of
magnetic clouds 1 and 2 (May 15, $\sim$ 09:00 UT and $\sim$ 21:00 UT, respectively).
}
 \label{fig-cartoon-Dt}
\end{figure}

Summarizing, we propose the following interpretation
of the sequence of events that occurred on
May 13, 2005: two ejective solar events occurred, which could be tracked
to the Earth environment and were observed as two attached,
but non-merged, magnetic clouds. Figure~\ref{fig-cartoon-Dt} shows the
position versus time (along the Sun-Earth line) of both interplanetary
events, as inferred from the sequence of the observations
used for our analyses from the Sun to 1~AU.
The presence of structures preserving multiple flux ropes,
as the multiple structure studied here, has been observed in some
cases \citep{Wang02b}.
Signatures of the interaction between ejecta, similar to those found in this
study, have also been identified from observations with the following
properties:
(i) acceleration of the leading ejecta and deceleration of the trailing one,
(ii) compressed field and plasma in the leading ejecta,
(iii) weak shock or disappearance of the shock driven by the second ejecta, and
(iv) strengthening of the shock driven by the leading one \citep{Farrugia04}.

 This study especially illustrates the need of combining solar, in situ and
remote sensing of interplanetary propagation of the ejecta, in order to reveal the
physical processes. Previous studies of the same event, but mostly limited to
one of these three domains have concluded on the presence of only one ejecta (with
varying characteristics depending on the study). In the present case, the peculiar
characteristics
of the in situ observed magnetic field first alert us on the
possible presence of two MCs.
We confirm this by performing a detailed modeling of the magnetic field exploring
various flux rope boundaries.
This stimulates a deeper study of data in the two other domains, with the new
perspective of testing two ejecta. The solar data were crucial to first confirm
the presence of two eruptions, and to define the timing and the spatial
organization
of the erupting magnetic configurations. This brought us to carry a deeper analysis
of interplanetary radio observations. Indeed, two type~II bursts,
traveling at
different velocities, were found.  These observations also permit to localize
where the interaction occurred (about half way between the Sun and the Earth). Then,
all the pieces of the puzzle finally fit together, re-enforcing
our interpretation separately in each domain.
However, the possibility of a single complex event
(as proposed by \citep{Yurchyshyn06,Reiner07}) cannot be fully discarded
and our proposed scenario, based on compelling evidence, must be further
investigated.

Such studies of interacting ejecta will greatly benefit from observations
taken with spacecraft in quadrature, as already done for some events
\citep[e.g., ][]{Rust05}, and as available with STEREO spacecrafts.
This has already be done with a spacecraft configuration far from
optimum conditions \citep{Harrison08}.
On top of providing in situ measurements
of physical parameters at three locations (combining ACE with STEREO) and
monitoring type~II bursts, there are imminent new possibilities of imaging ICMEs
in their journey through interplanetary space.
Then, we will have the opportunity to derive precise
constraints on
the physical mechanisms of interaction of an ejecta with a faster one, or
simply with a fast overtaking solar wind.

\begin{acknowledgements}
The authors would like to thank to the International Space Science
Institute (ISSI, Bern, Switzerland) for supporting the project
'The Stages of Sun-Earth Connection', lead by C. Cid.
C.C, Y.C, and E.S. acknowledge financial support from the Comisi\'on
Interministerial de Ciencia y Tecnología(CICYT) of Spain
(ESP 2005-07290-C02-01 and ESP 2006-08459).
S.D. thanks the Argentinean grant UBACyT X425.
S.D. and C.H.M. thank the Argentinean grants:
UBACyT X329 and PICT 03-33370 (ANPCyT).
H.C., S.D., and C.H.M. are members of the Carrera del
Investigador Cient\'{i}fico, CONICET.
B.S., P.D., and S.P. are doing partly this project in the frame of the
European Network ``SOLAIRE''.
C.H.M., B.S. and P.D. acknowledge financial support from CNRS (France) and
CONICET (Argentina) through their cooperative science program (N$^o$ 20326).
L.R. and A.N.Z. acknowledge support from the Belgian Federal Science Policy
Office through the ESA-PRODEX programme.
The work at the University of Barcelona was supported by the Ministerio
de Educaci\'on y Ciencia under the projects AYA2004-0322 and AYA2007-60724.
We are grateful to the ACE/SWEPAM and ACE/MAG teams, for the data used for
this work.
We would also like to thank OmniWeb from where we downloaded the
preliminar Dst index.
The CME catalog is generated and maintained at the CDAW Data Center by NASA
and The Catholic University of America in cooperation with the
Naval Research Laboratory.
We acknowledge the use of Wind WAVES/TNR radio data.
The Optical Solar PAtrol Network (OSPAN) project, previously called
ISOON, is a collaboration between the Air Force Research Laboratory Space
Vehicles Directorate and the National Solar Observatory.
We acknowledge the use of TRACE data.
EIT, LASCO and MDI data are a courtesy of SOHO/EIT,
SOHO/LASCO and SOHO/MDI consortia. SOHO is a project of
international cooperation between ESA and NASA.
The authors thank S. Hoang for providing clear informations
on Wind/WAVES and the referees for their helpful suggestions.
\end{acknowledgements}


\bibliographystyle{agufull}
\bibliography{mc}
\IfFileExists{\jobname.bbl}{}
{\typeout{}
\typeout{****************************************************}
\typeout{****************************************************}
\typeout{** Please run "bibtex \jobname" to optain}
\typeout{** the bibliography and then re-run LaTeX}
\typeout{** twice to fix the references!}
\typeout{****************************************************}
\typeout{****************************************************}
\typeout{}
}

\end{article}
\end{document}